\documentclass[a4paper,11pt]{article}

\usepackage[]{a4}
\usepackage{amsmath}
\usepackage{amsfonts}
\usepackage{amssymb}
\usepackage{graphicx}

\usepackage{geometry}

\usepackage{authblk}

\DeclareMathOperator{\sgn}{sgn}

\begin{document}

\title{Implications for constrained supersymmetry of combined H.E.S.S.
  observations of dwarf galaxies, the Galactic halo and the Galactic
  Centre}

\author[1]{Joachim Ripken} 
\author[1]{Jan Conrad} 
\author[2]{Pat Scott}
\affil[1]{Oskar Klein Centre for Cosmoparticle Physics and Department
  of Physics, Stockholm University, AlbaNova University Centre,
  SE-10691 Stockholm, Sweden} 
\affil[2]{Department of Physics, McGill
  University, 3600 rue University, Montr\'eal, QC, H3A 2T8, Canada}

\date{}
\maketitle

\begin{abstract}
  In order to place limits on dark matter (DM) properties using
  $\gamma$-ray observations, previous analyses have often assumed a
  very simple parametrisation of the $\gamma$-ray annihilation yield;
  typically, it has been assumed that annihilation proceeds through a
  single channel only.  In realistic DM models, annihilation may occur
  into many different final states, making this quite a rough ansatz.
  With additional processes like virtual internal bremsstrahlung and
  final state radiation, this ansatz becomes even more incorrect, and
  the need for scans of explicit model parameter spaces becomes clear.
  Here we present scans of the parameter space of the Constrained
  Minimal Supersymmetric Standard Model (CMSSM), considering
  $\gamma$-ray spectra from three dwarf galaxies, the Galactic Centre
  region and the broader Galactic halo, as obtained with the
  High-Energy Stereoscopic System (H.E.S.S.).  We present a series of
  likelihood scans combining the H.E.S.S. data with other experimental
  results. We show that observations of the Sagittarius, Carina and
  Sculptor dwarf galaxies disfavour the coannihilation region of the
  CMSSM and models with large annihilation cross-sections.  This is
  true even under reasonable assumptions about the DM density
  profiles, and constitutes the strongest constraint to date on
  coannihilation models within the CMSSM.  The Galactic halo has a
  similar, but weaker, effect.  The Galactic Centre search is
  complicated by a strong (unknown) $\gamma$-ray source, and we see no
  effect on the CMSSM parameter space when assuming a realistic
  Galactic Centre DM density profile.
\end{abstract} 

\section{Introduction}
The nature of the dark matter (DM) is still unclear, because its
existence is so far only evident via its gravitational interaction
with normal matter.

There are many possible ways to find DM, other than through is
gravitational interactions. Direct detection experiments try to
observe nuclear recoils from weak interactions with DM particles.
Accelerator experiments search for hints of physics beyond the
\textit{standard model} (SM) of particle physics, which may provide
clues as to the identity of dark matter. Indirect detection
experiments try to identify secondary products of DM annihilation or
decay, such as photons, neutrinos and anti-particles.  Experiments
typically search for characteristic spectral signatures of DM in the
cosmic fluxes of such particles, allowing them to (hopefully)
differentiate the DM signal from the myriad of astrophysical
backgrounds they face.

There are many candidates for DM in extensions of the SM. The most
popular is the neutralino, a linear combination of the superpartners
of the neutral Higgs and electroweak gauge bosons seen in
supersymmetric (SUSY) extensions of the SM.  If $R$-parity is
conserved, and the lightest neutralino is also the lightest
supersymmetric particle (LSP), it can -- depending on the underlying
SUSY model parameters -- deliver a relic density in the favoured range
$0.094 \leq \Omega h^{2} \leq 0.129$. Neutralinos are also Majorana
particles, so would self-annihilate.  If SUSY is to constitute a valid
solution to the well-known \textit{hierarchy problem} of the SM, it
must be broken at $\sim$1\,TeV, giving sparticles such as the lightest
neutralino masses of between $\sim$10\,GeV and $\sim$10\,TeV.

In the annihilation process, very high energy (VHE) $\gamma$-ray
photons are produced with energies up to the neutralino mass. The
emissivity of annihilating DM is proportional to $\varrho^{2}$, the
square of the DM density. It is thus useful to search for VHE
$\gamma$-radiation from regions where a high density of DM is
expected. One such region is the centre of our own galaxy, the
Galactic Centre (GC).

Limits on DM annihilation are generally based on assumptions about the
form of the annihilation spectrum, ignoring the individual spectra of
actual SUSY (or any other) models. This was perhaps reasonable until
it was found that internal bremsstrahlung (IB), consisting of both
final-state radiation (FSR) and virtual IB (VIB), can make large
contributions to the photon spectrum \cite{bib_IB}.  In this case,
gamma-ray spectra from different supersymmetric models can be very
different, even when the neutralino mass is kept fixed. With this new
development, it is necessary to compare the observed and predicted
energy spectra from annihilation processes on an individual,
model-by-model basis.  This was first performed in a full SUSY scan
using \textit{Fermi}-LAT data on the dwarf galaxy Segue 1
\cite{bib_Pat}.

The GC region has also been observed by the High Energy Stereoscopic
System (H.E.S.S.), and high-energy gamma radiation has been detected
\cite{bib_hessGC1, bib_hessGC5}.  Because the observations seem to be
incompatible with the total observed flux coming exclusively from
neutralino annihilation, the hypotheses that DM annihilation makes a
subdominant contribution has been investigated, resulting in limits on
the DM self-annihilation cross-section \cite{bib_hessGC2, bib_hessGC4}.

In this article we show the results of two full model scans in the
parameter space of the Constrained Minimal Supersymmetric SM (CMSSM),
comparing model predictions with H.E.S.S. data from the Sagittarius
(SgrD), Carina and Sculptor dwarf galaxies, as well as the Galactic
halo and Galactic Centre.  First we show a simple random scan,
producing a set of CMSSM models compatible with constrains on the
relic density and accelerator bounds included in DarkSUSY 5.0.4
\cite{bib_DS, bib_DSweb}.  Later we show more advanced statistical
scans, using the SuperBayeS package \cite{bib_SB1, bib_SB4, bib_SBweb,
  bib_SB3, bib_SB5, bib_SB2, bib_SB6}.

In section \ref{sec_hess} we introduce the H.E.S.S. experiment and the
data that we use for this work. Section \ref{sec_theo} is about the
theoretical framework of supersymmetric DM, and Section \ref{sec_ana}
describes our analysis of the H.E.S.S. data. Section
\ref{subsec_GC1} gives our results for the random scan using a
spectrum from the GC source. Section \ref{subsec_GC2} introduces
the CMSSM parameter scan with SuperBayeS, considering the same GC
spectrum. Section \ref{subsec_SgD} describes a SuperBayeS scan
taking into account the H.E.S.S. observations on the SgrD, whilst
Secs.~\ref{sec_2dwarfs} and \ref{sec_halo} introduce further
constraints from the Carina and Sculptor dwarfs, and the Galactic
halo, respectively. Section \ref{sec_sum} finishes with a summary and
outlook.

\section{The H.E.S.S. telescope and data} \label{sec_hess}

H.E.S.S. is a system of $4$ imaging atmospheric \v{C}erenkov
telescopes located in the Khomas highlands of Namibia, $120 \,
\text{km}$ south west of Windhoek, and $1800 \, \text{m}$ above sea
level. It is a $\gamma$-ray observatory sensitive to photons with
energies between around $100 \, \text{GeV}$ and $100 \, \text{TeV}$.
The energy resolution is better than $15 \%$. The angular resolution
is better than $0.1 ^{\circ}$ per event. \cite{bib_hesscrab}

The observed $\gamma$-ray spectrum for our scans including the GC data
is from \cite{bib_hessGC5}. It contains $92.9 \, \text{h}$ of
observations in the years $2004$, $2005$ and $2006$. For the following
analysis we employed the spectral points seen in the left-hand
subfigure of Figure 2 in \cite{bib_hessGC5}. These data were already
deconvolved from the instrumental response at the time of publication
(see Ref.~\cite{bib_hessGC5} for details), removing any need for us to
convolve our predicted CMSSM spectra with the H.E.S.S. response.

H.E.S.S. observed the SgrD in June 2006 for $\sim$$12\,\text{h}$. No
significant $\gamma$-ray excess was found and a flux upper limit of
$\Phi (E > 250 \, \text{GeV}) = 3.6 \cdot 10^{-12} \,
\text{cm}^{-2}\,\text{s}^{-1}$ (95\% \text{CL}) was calculated. Using
these observations, and assuming a generic annihilation spectrum as
well as two different DM density profiles, upper limits on the
annihilation cross section $\langle \sigma v \rangle$ as function of
the neutralino mass $m_{\chi}$ were calculated \cite{bib_hessSGD}.

Observations of the Carina and Sculptor dwarf galaxies took place
between January $2008$ and December $2009$ with $\sim$$15 \, \text{h}$
on Carina and $\sim$$12 \, \text{h}$ on Sculptor. Also here no
significant $\gamma$-ray excess was found leading also to upper limits
on the annihilation cross section as function of the neutralino mass
\cite{bib_carscul}.

Observations of the region around the GC were also used to search for
diffuse $\gamma$-radiation originating from DM annihilation in the
galactic halo. This radiation has not been found, so that again upper
limits were calculated \cite{bib_halo}.

\section{Theoretical framework} \label{sec_theo}

Adding the minimal additional particle content required to
supersymmetrise the SM, along with the most general `soft'
SUSY-breaking Lagrangian terms (required to break but retain SUSY as a
solution of the hierarchy problem), one arrives at the Minimal
Supersymmetric Standard Model (MSSM).  The addition of the soft terms
introduces over 100 new parameters to the model, so even in the MSSM,
simplifying assumptions are required in order to make any meaningful
estimates of the parameters of the model.  One way to arrive at such a
simplified version of the model is to choose a specific breaking
scheme, with the symmetry breaking parameters set at a high energy
scale, and then use renormalisation group equations to arrive at the
corresponding masses and couplings at lower energies. One particular
example, which we will consider in this paper, is the CMSSM, where the
model is defined
by five free parameters:\\
\begin{equation}
m_0;m_\frac12 ;A_0; \text{tan} \beta; \text{sgn} \mu;
\end{equation}
Here $m_0$ is the universal scalar mass, $m_\frac12$ the gaugino mass
parameter, $A_0$ the trilinear coupling between Higgs bosons, squarks
and sleptons, $\tan\beta$ the ratio of vacuum expectation values of
up-type and down-type Higgs bosons, and $\sgn\mu$ the sign of the
Higgs mixing parameter.  The parameters $m_0$, $m_\frac12$ and $A_0$
are defined at the GUT scale (10$^{16}$ GeV), whereas $\tan\beta$ and
$\sgn\mu$ are defined at the weak scale.  Most authors define the CMSSM
 and mSUGRA (a `minimal SUperGRAvity-inspired' parametrisation of
the MSSM) identically, and refer to them interchangeably; some other
definitions of mSUGRA do exist, but the CMSSM is unambiguous.

In the literature, several regions have been identified where a
neutralino LSP provides the right relic abundance of dark matter.
These regions are then further constrained by accelerator searches.
The regions that are still viable are the stau coannihilation region,
where the stau is almost degenerate with the LSP (and the correct DM
abundance is achieved by coannihilations), the focus point region
(where the LSP is Higgsino-like), and the funnel regions, where LSP
annihilation is increased by resonance interactions with MSSM Higgs
particles.

\section{Analysis} \label{sec_ana}

The flux delivered by annihilating DM can be calculated with
\cite{bib_dmgamma}:
\begin{equation}
\begin{split}
\label{flux}
\Phi(E) &= 2.8 \cdot 10^{-12} \, \text{cm}^{-2} \text{s}^{-1}
\text{sr}^{-1} \cdot \frac{dN_{\gamma}}{dE} \frac{\langle \sigma v
  \rangle}{\text{pb} \cdot c} \Bigl{(} \frac{1 \,
  \text{TeV}}{m_{\chi}} \Bigr{)}^{2}
\cdot \bar{J}(\Delta \Omega)\Delta \Omega \\
\Delta \Omega &= \frac{1}{8.5 \text{kpc} \cdot
  (0.3 \, \text{GeV} \, \text{cm}^{-3})^{2}} \int_{\Delta \Omega} d
\Omega \int_{\text{los}} ds \, \varrho^{2}
\end{split}
\end{equation} \label{eq_dmflux}
where $dN/dE$ describes the photon spectrum per annihilation, $\langle
\sigma v \rangle$ is the thermally-averaged, velocity-weighted annihilation 
cross-section in the zero velocity limit (in the following simply 
denoted ``cross section''), $m_{\chi}$ is the mass of the annihilating 
DM particle and
$\varrho$ is its density, which is integrated along the line of sight
(los) and over the observed solid angle of 
$\Delta \Omega = 1.16 \cdot 10^{-5} \, \text{sr}$ for the GC, 
$\Delta \Omega = 2 \cdot 10^{-5} \, \text{sr}$ for SgrD, and $\Delta 
\Omega = 10^{-5} \, \text{sr}$ for Carina and Sculptor. For the galactic 
halo the signal- and background regions are defined more complicated than 
for the other targets. The $J$-factor for this $\bar{J}(\Delta \Omega)$ 
represents the difference between the averaged line of sight integral 
in the signal and in the background region.

\section{CMSSM random scan with data from the Galactic Centre}
\label{subsec_GC1}

\begin{figure}[ht] 
\begin{center}
\includegraphics[width=0.44\linewidth]{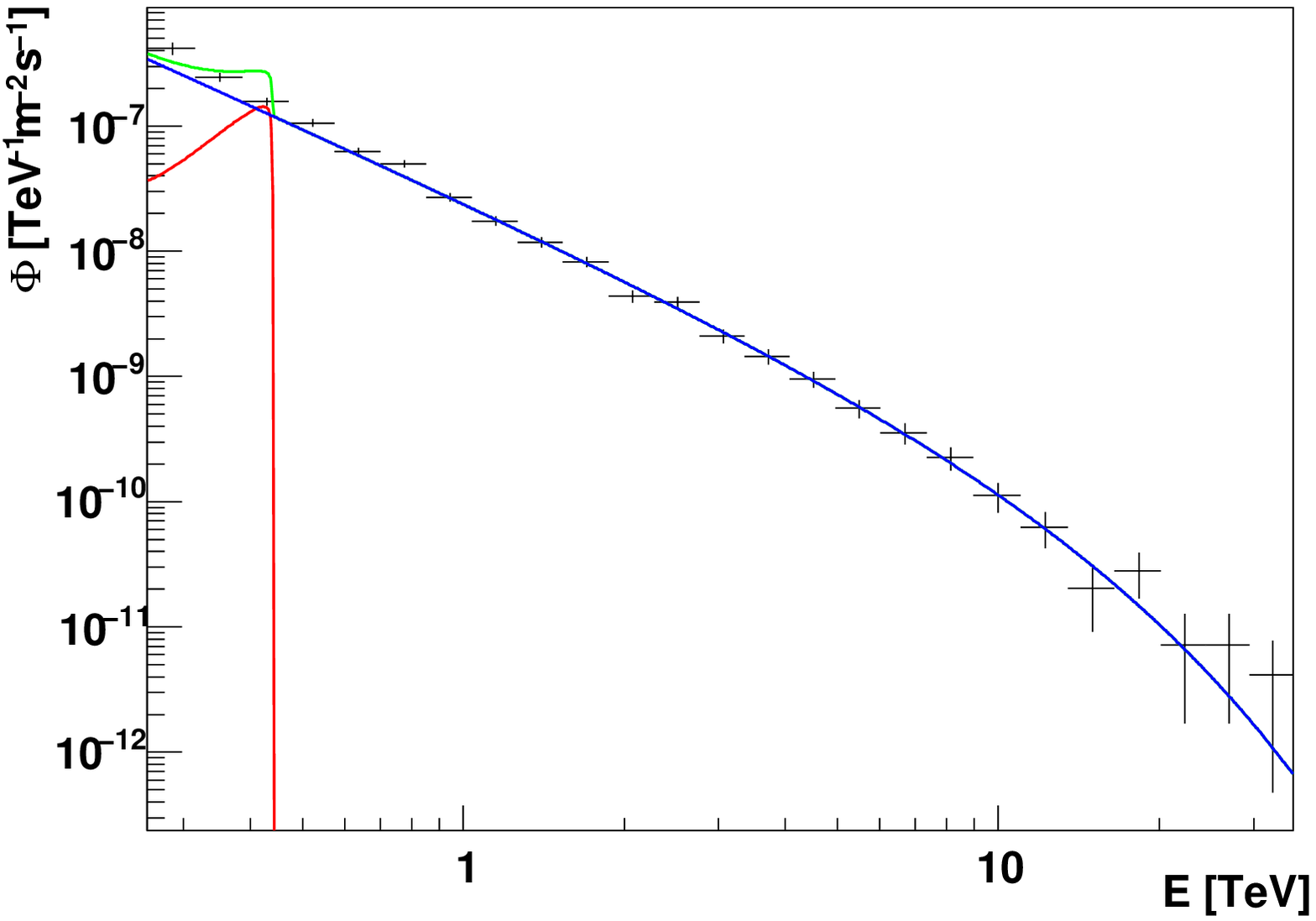}
\includegraphics[width=0.44\linewidth]{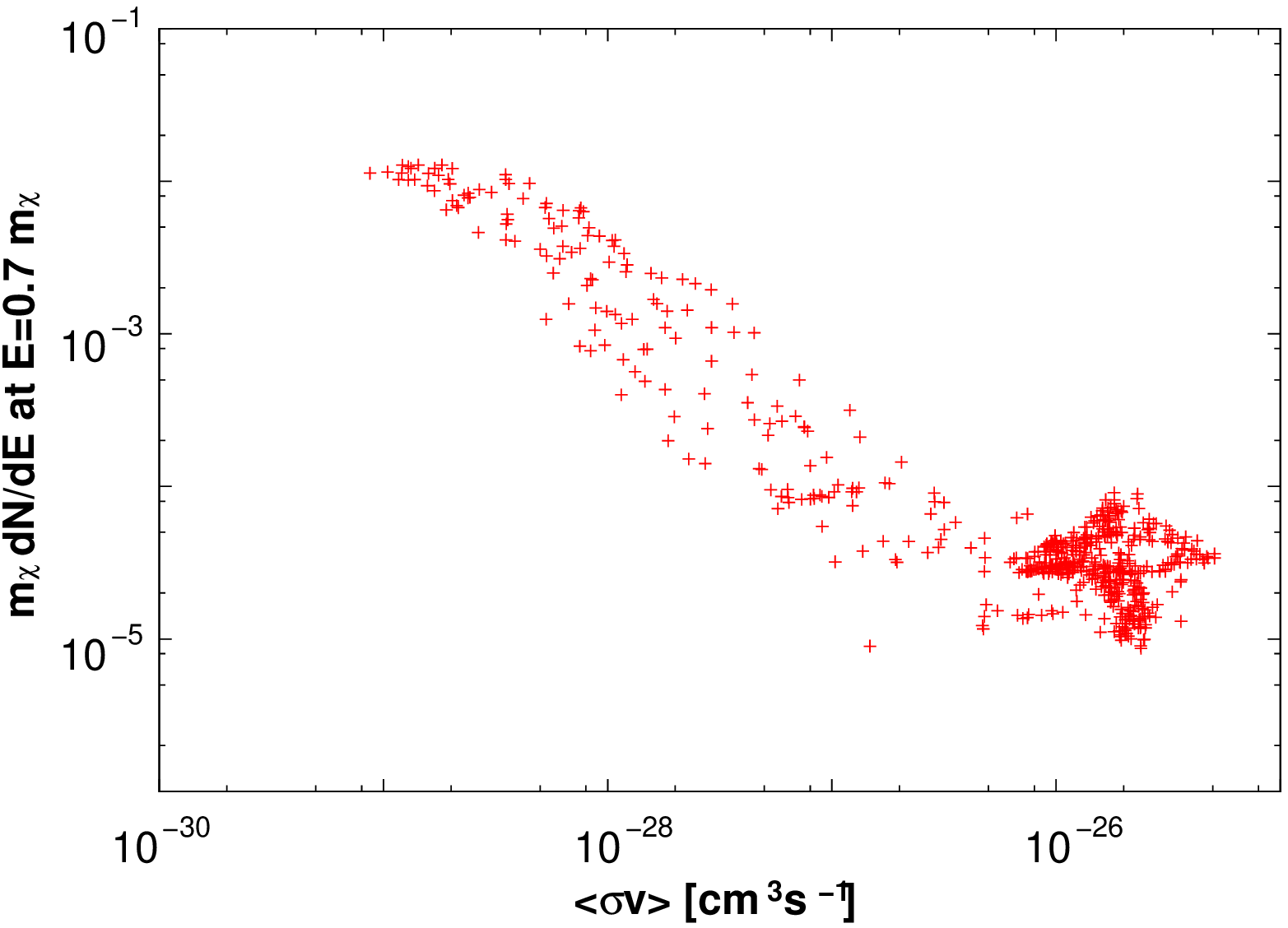}
\end{center} 
\caption{\textbf{Left}: Example of a comparison between data and model
  annihilation spectrum. The crosses show the measured spectrum
  published by the H.E.S.S. collaboration \cite{bib_hessGC5}. The red
  line shows $\Phi_\text{DM}$ the calculated dark matter spectrum for
  $m_{0} = 438 \, \text{GeV}$, $m_\frac12 = 1030 \, \text{GeV}$,
  $A_{0} = 0$, $\tan \beta = 39.1$, $\sgn \mu = +1$ and
  $\bar{J}(\Delta \Omega) \Delta \Omega = 350 \, \text{sr}$ (Moore
  profile). The blue line represents $\Phi_{\text{bg}}$, the
  background model (a power law with exponential cutoff) that delivers
  the best fit as part of $\Phi_{\text{total}} = \Phi_{\text{DM}} +
  \Phi_{\text{bg}}$. \textbf{Right}: Correlation plot for the
  annihilation cross section $\langle \sigma v \rangle$ and
  $\gamma$-ray yield $dN/dx$ with $x = E_{\gamma}/m_{\chi}$ at $x =
  0.7$, showing the indicative number of photons per annihilation with
  energies just below the WIMP mass.  Because IB has a harder
  gamma-ray spectrum than pion decay, for a fixed $\langle \sigma v
  \rangle$ models with larger yields at $E=0.7m_\chi$ show stronger
  IB. Here we see that for the points that passed our relic density
  and accelerator cuts, the yield into photons with energies near the
  WIMP mass decreases as the cross-section increases, indicating that
  IB is much stronger in models with lower annihilation
  cross-sections.}
\label{fig_model}
\end{figure} 

\begin{figure}[t!] 
\begin{center}
\includegraphics[width=0.9\linewidth]{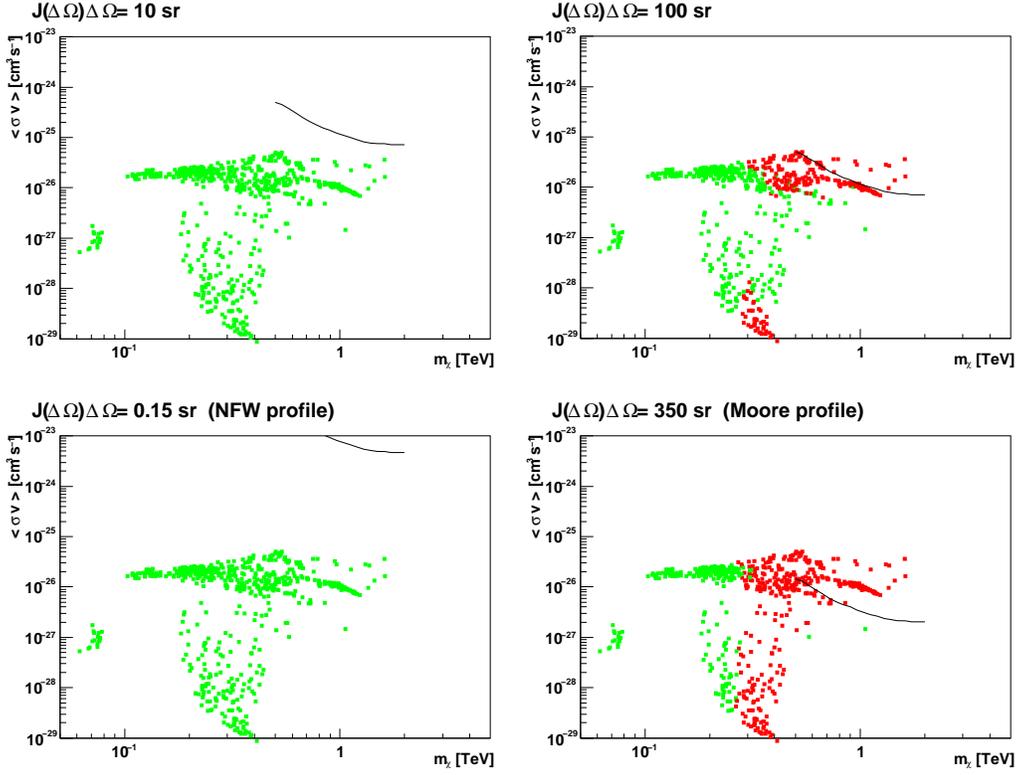} 
\end{center} 
\caption{Rejection plots in the plane spanned by the neutralino mass
  $m_{\chi}$ and the annihilation cross section $\langle \sigma v
  \rangle$. The green points represent models consistent with the
  data, and red points models that are not consistent (at $90\% \,
  \text{CL}$). The black line shows the upper limit of $\langle \sigma
  v \rangle$ as function of $m_{\chi}$, if annihilation of neutralinos
  proceeded entirely as $\chi \, \chi \rightarrow b \, \bar{b}$. In
  the top row, $\bar{J}(\Delta \Omega)\Delta \Omega$ increases from
  $10 \, \text{sr}$ (upper left) to $100 \, \text{sr}$ (upper right).
  The plots for an NFW profile ($\bar{J}(\Delta \Omega)\Delta \Omega =
  0.15 \, \text{sr}$) and a Moore profile ($\bar{J}(\Delta
  \Omega)\Delta \Omega) = 350 \, \text{sr}$)
  are shown in the lower row.}
\label{fig_rej}
\end{figure}

Earlier analysis of the GC by H.E.S.S. showed that DM alone cannot be
responsible for the observed spectrum \cite{bib_hessGC2}.  We
therefore consider an energy spectrum
\begin{equation}
\Phi_{\text{total}}(E) = \Phi_{\text{DM}}(E) + \Phi_{\text{bg}}(E),
\end{equation}
composed of a DM component $\Phi_{\text{DM}}$ and an
empirically-determined background, assumed to take the form of a power
law with an exponential cut-off:
\begin{equation}
  \Phi_{\text{bg}}(E) = \Phi_{0} \cdot 
  \Bigl{(}\frac{E}{1 \, \text{TeV}}\Bigr{)}^{-\Gamma} \exp(-E/E_\text{cut}),
\end{equation}
where $\Phi_{0}$ is the flux normalisation, $\Gamma$ represents the
spectral slope and $E_\text{cut}$ the cutoff energy.  Such a form for
the background provides quite a good fit to the observed spectrum
\cite{bib_hessGC2, bib_hessGC5}, and is generally representative of
typical astrophysical gamma-ray sources.  In our scans, we fit
$\Phi_{0}$, $\Gamma$ and $E_\text{cut}$ individually for each DM model
in the CMSSM.

As a first check, we randomly chose $622$ CMSSM models whose relic
densities fit within the observed band ($0.094 \leq \Omega_{\text{DM}}
h^{2} \leq 0.129$) from $3$ years WMAP observations \cite{bib_WMAP3},
and pass similar accelerator bounds included in DarkSUSY 5.0.4. The
CMSSM parameters in this scan lie in the following ranges: $10 \,
\text{GeV} \leq m_{0} \leq 1000 \, \text{GeV}$, $10 \, \text{GeV} \leq
m_\frac12 \leq 1000 \, \text{GeV}$, $A_{0} = 0$, $0 \leq \tan \beta
\leq 60$, $\sgn \mu \in \{-1, 1\}$.  Whether a model (CMSSM parameters
and chosen $J$ factor) is compatible with the measured data is decided
by a $\chi^{2}$-test. We fit $\Phi_{\text{total}}(E)$ to the data for
each model, keeping the parameters for $\Phi_{\text{DM}}(E)$ fixed and
the parameters for $\Phi_{\text{bg}}(E)$ free.  An example of such a
comparison is shown in the left panel of Figure \ref{fig_model}.

A model (with an assumed value for $\bar{J}(\Delta \Omega)\Delta
\Omega$) is defined as compatible with the data if the resulting
$\chi^{2} < 14.04$; the $90\%$ threshold value of the $\chi^{2}$
distribution with $N_{\text{bins}}-N_{\text{free}} = 25-3$ degrees of
freedom.  Results can be seen in Figure \ref{fig_rej}.  Here we show
whether a model -- represented by a point in the $m_{\chi}$-$\langle
\sigma v \rangle$-plane -- is compatible with the measured spectrum or
not, given different assumed values of $\bar{J}(\Delta \Omega)\Delta
\Omega$. We also indicate the upper limit obtained if one assumes
100\% annihilation into $b \, \bar{b}$, as has often been done in
previous analysis. For comparison, the $J$ factor for an NFW profile
would be $\bar{J}(\Delta \Omega)\Delta \Omega \lvert_{\text{NFW}} =
0.15 \, \text{sr}$ and for a Moore profile $\bar{J}(\Delta
\Omega)\Delta \Omega \lvert_{\text{Moore}} = 350 \, \text{sr}$.

For $\bar{J}(\Delta \Omega) \Delta \Omega \gtrsim 10 \, \text{sr}$ the
data begin to limit models from high cross-sections downward (into the
focus point region). In addition, the parameter space is truncated
from low cross-sections upward (into the coannihilation region) due to
IB, as the number of photons from these processes and the annihilation
cross-section are anti-correlated (see the right panel of Figure
\ref{fig_model}). For $\bar{J}(\Delta \Omega) \Delta \Omega \gtrsim
100$ the two limiting fronts meet. Models with $m_{\chi} \lesssim 200
\, \text{GeV}$ remain allowed, because they do not affect the spectrum
in the energy range observed by H.E.S.S.. A few models with $m_{\chi}
\gtrsim 500 \, \text{GeV} - 1000 \, \text{GeV}$ and $\langle \sigma v
\rangle \lesssim 10^{-27} \, \text{cm}^{-3} \text{s}^{-1}$ also remain
allowed.

A random scan is however not sufficient when dealing with a
complicated parameter space with many dimensions, such as the CMSSM.
Nuisance parameters that could substantially affect model predictions,
such as the top quark mass, should be taken into account.  Points
should be measured against a whole range of observables, and given a
properly-defined statistical likelihood rather than just ruled in or
out.  Sophisticated scanning algorithms should be used to make sure
that all relevant parts of the parameter space have been probed, and
in a way that allows valid statistical inference to be performed on
the resultant points.  For these reasons, we performed additional
likelihood-based scans.

\section{CMSSM likelihood scan with data from the Galactic Centre} 
\label{subsec_GC2}

\begin{figure}[ht]  
\begin{center}
\includegraphics[width=0.31\linewidth]{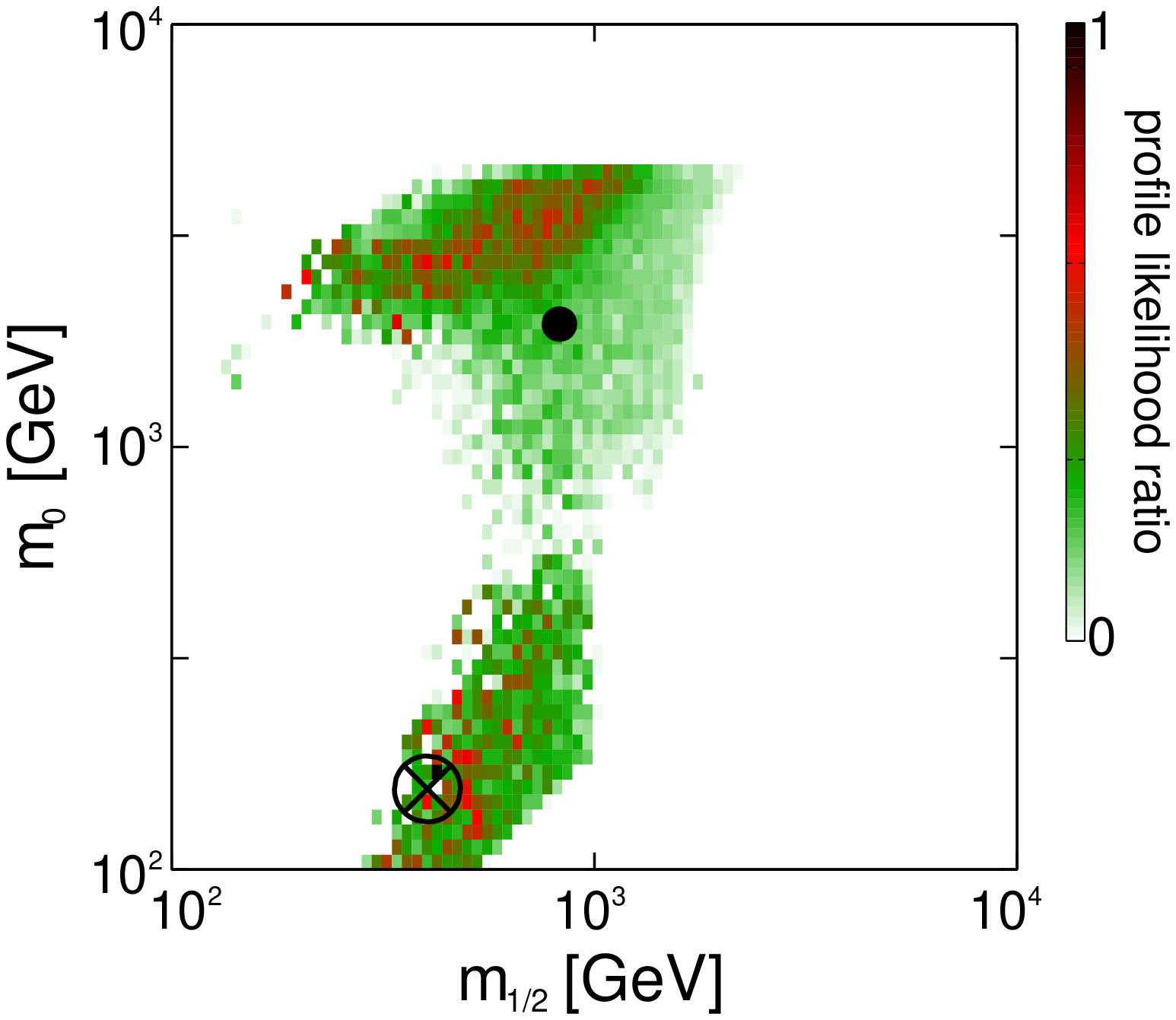}
\includegraphics[width=0.31\linewidth]{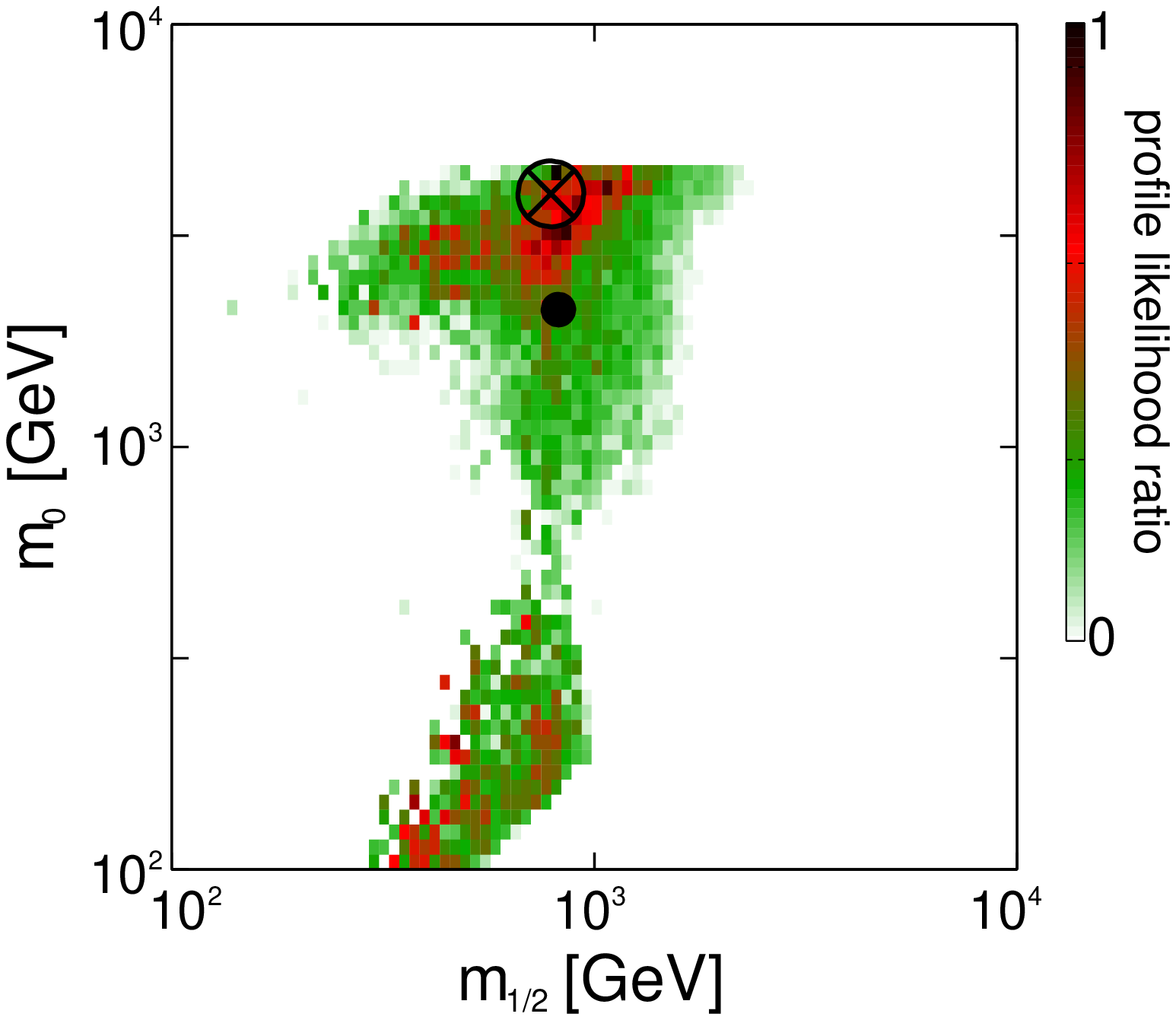}
\includegraphics[width=0.31\linewidth]{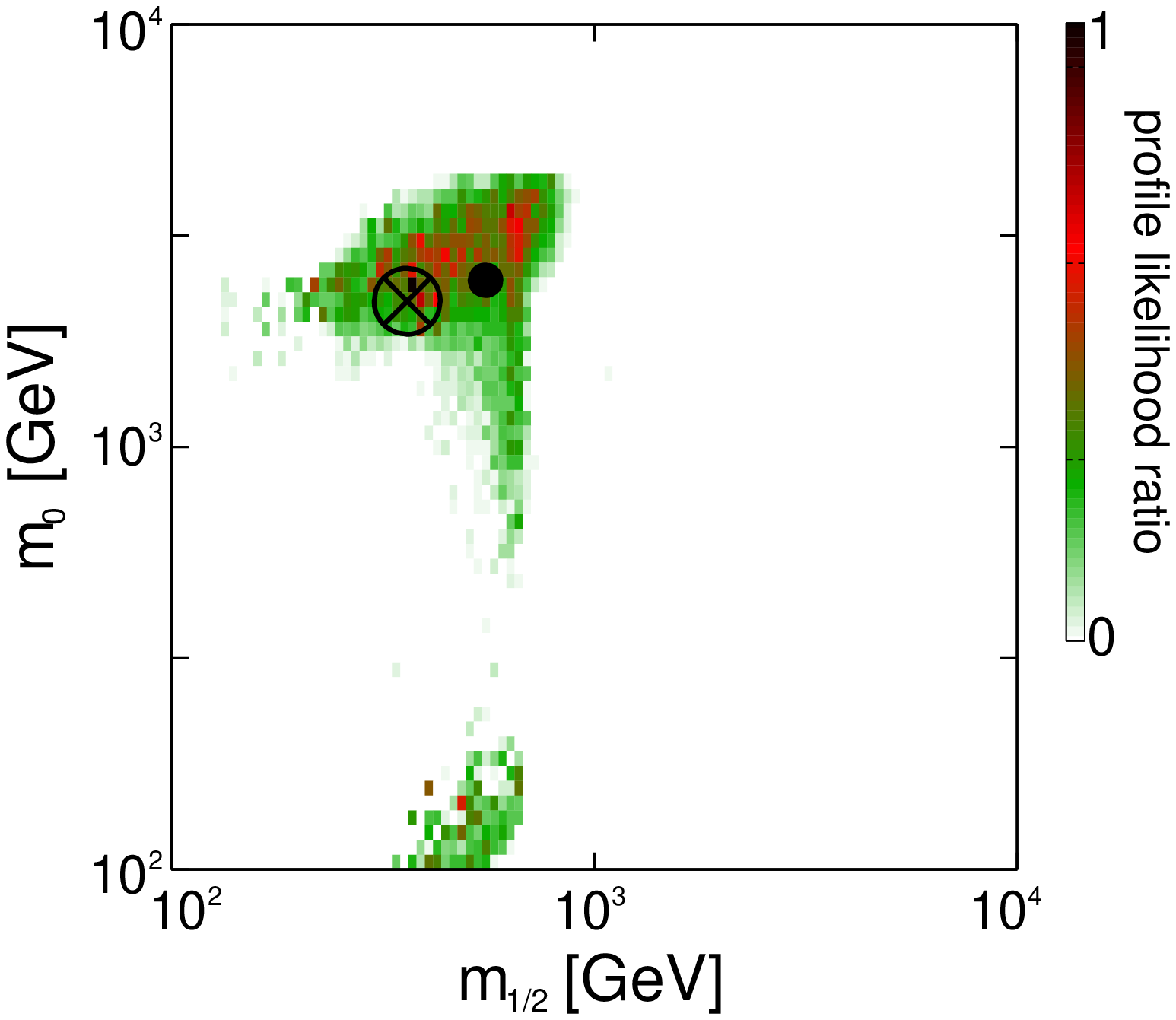}
\includegraphics[width=0.31\linewidth]{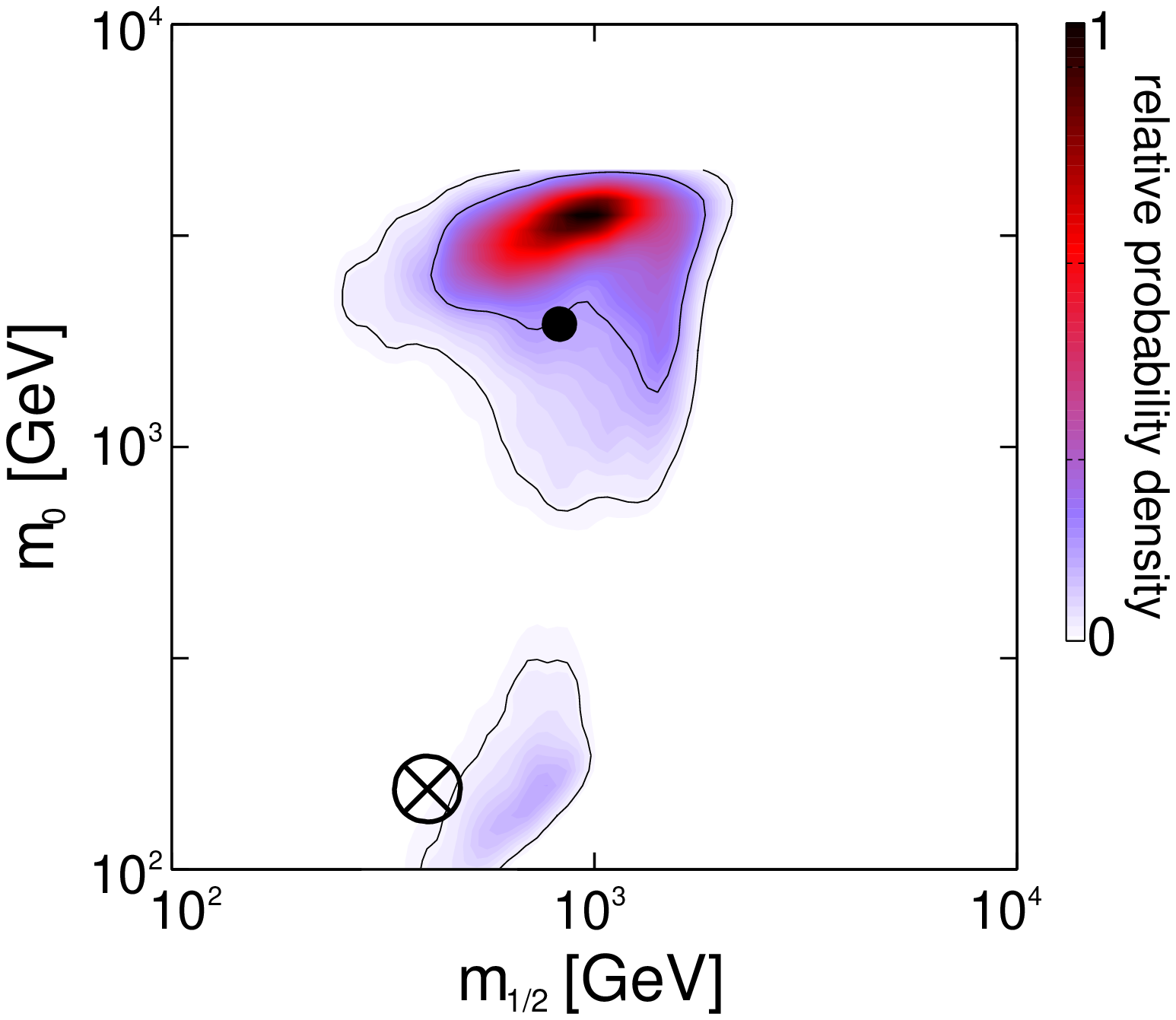}
\includegraphics[width=0.31\linewidth]{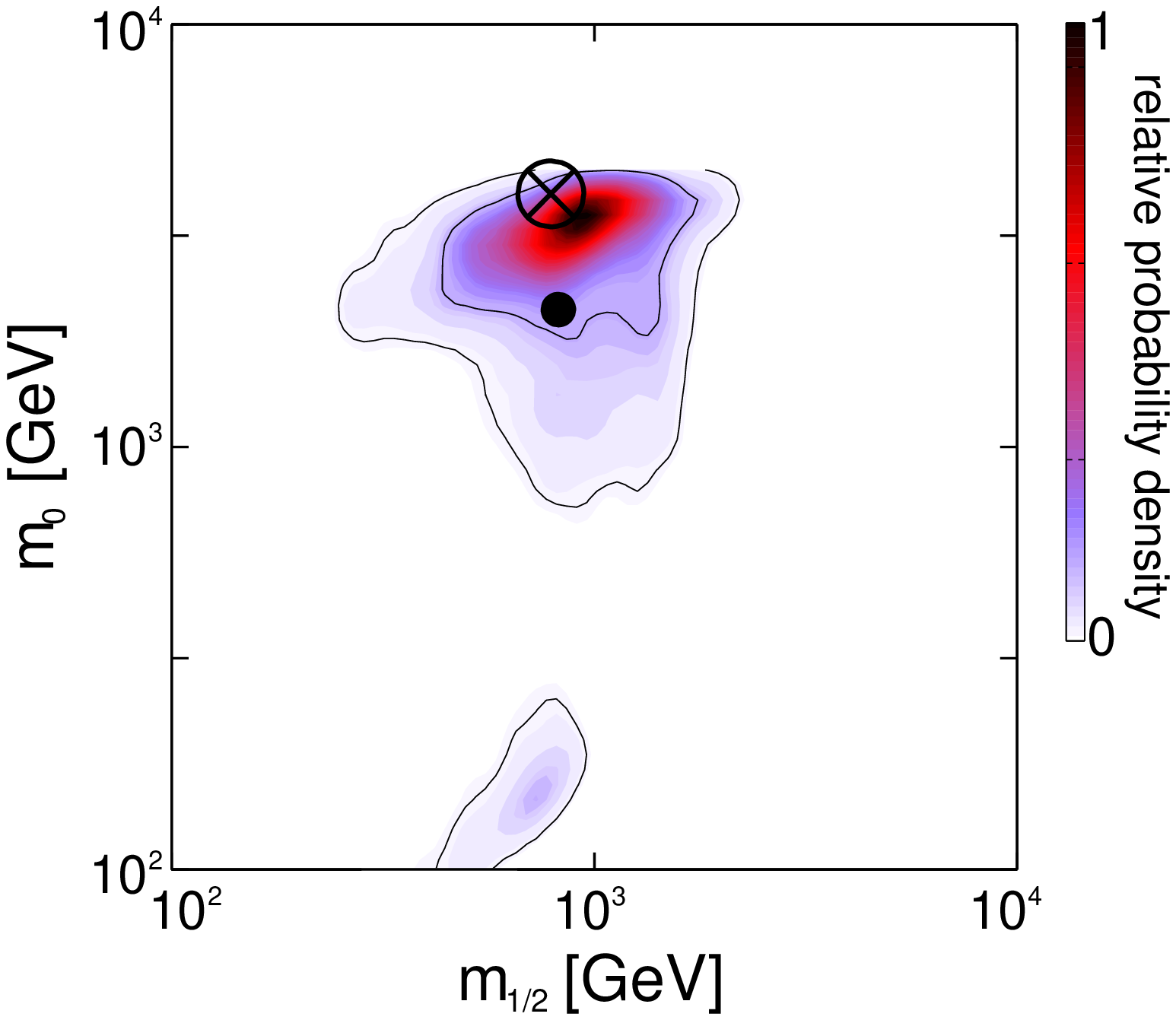}
\includegraphics[width=0.31\linewidth]{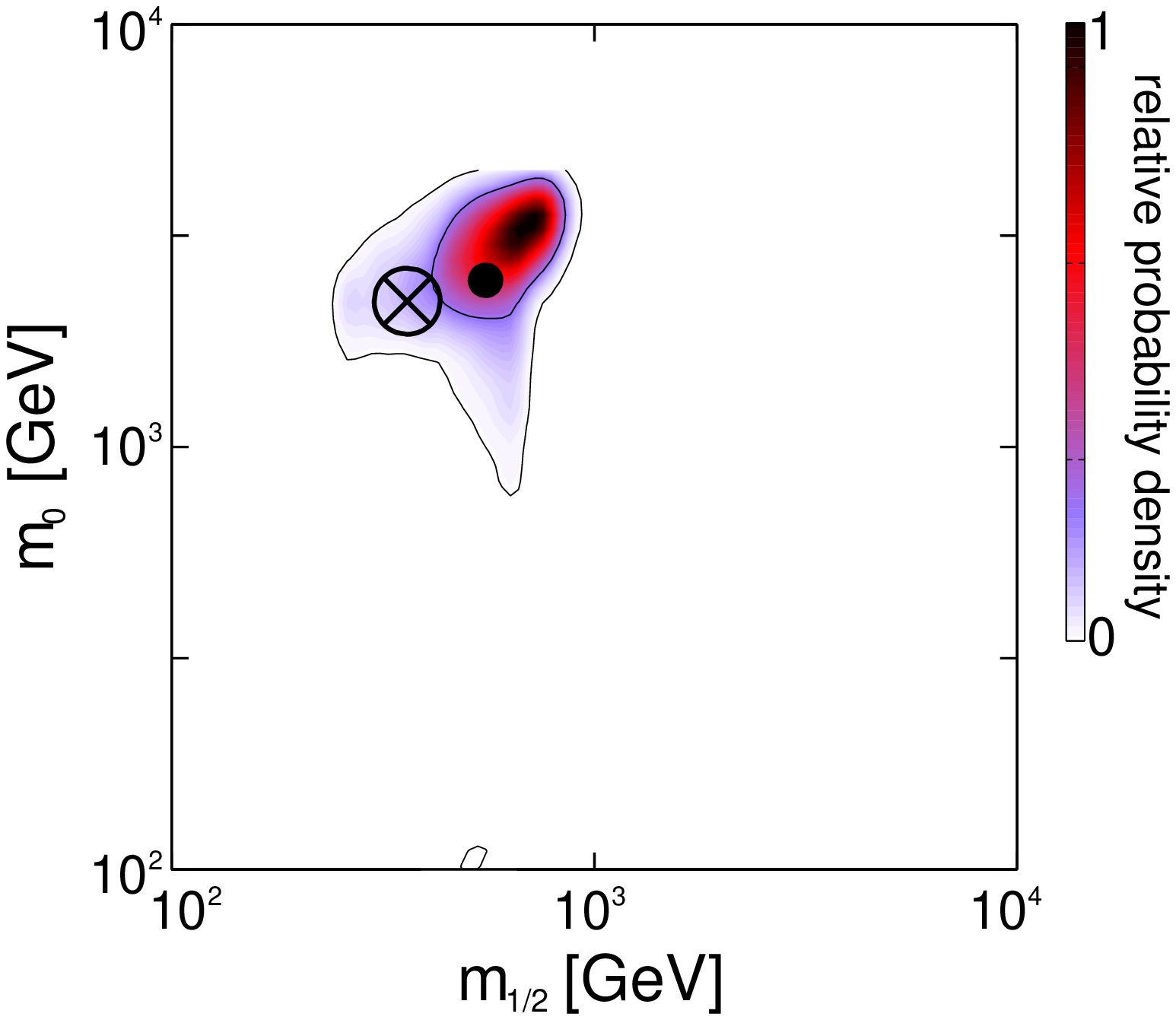}
\end{center} 
\caption{Profile likelihood ratios
  $\mathcal{L}/\mathcal{L}_{\text{best fit}}$ (upper row) and
  posterior probability density functions (lower row) normalized to
  the best fit point in the $m_\frac12$-$m_{0}$ plane. In the left
  column H.E.S.S.  data from the Galactic Centre have not been
  included in the scans.  For the other plots, $\bar{J}(\Delta
  \Omega)\Delta \Omega = 10 \, \text{sr}$ (middle) and $100 \,
  \text{sr}$ (right).  The $\otimes$ marks the best fit point, while
  the $\bullet$ marks the centre of gravity of the distribution.
  Contours in the lower plots surround $68 \%$ and $95 \%$ credible
  regions.}
\label{fig_sb1}
\end{figure}

\begin{figure}[ht]  
\begin{center}
\includegraphics[width=0.31\linewidth]{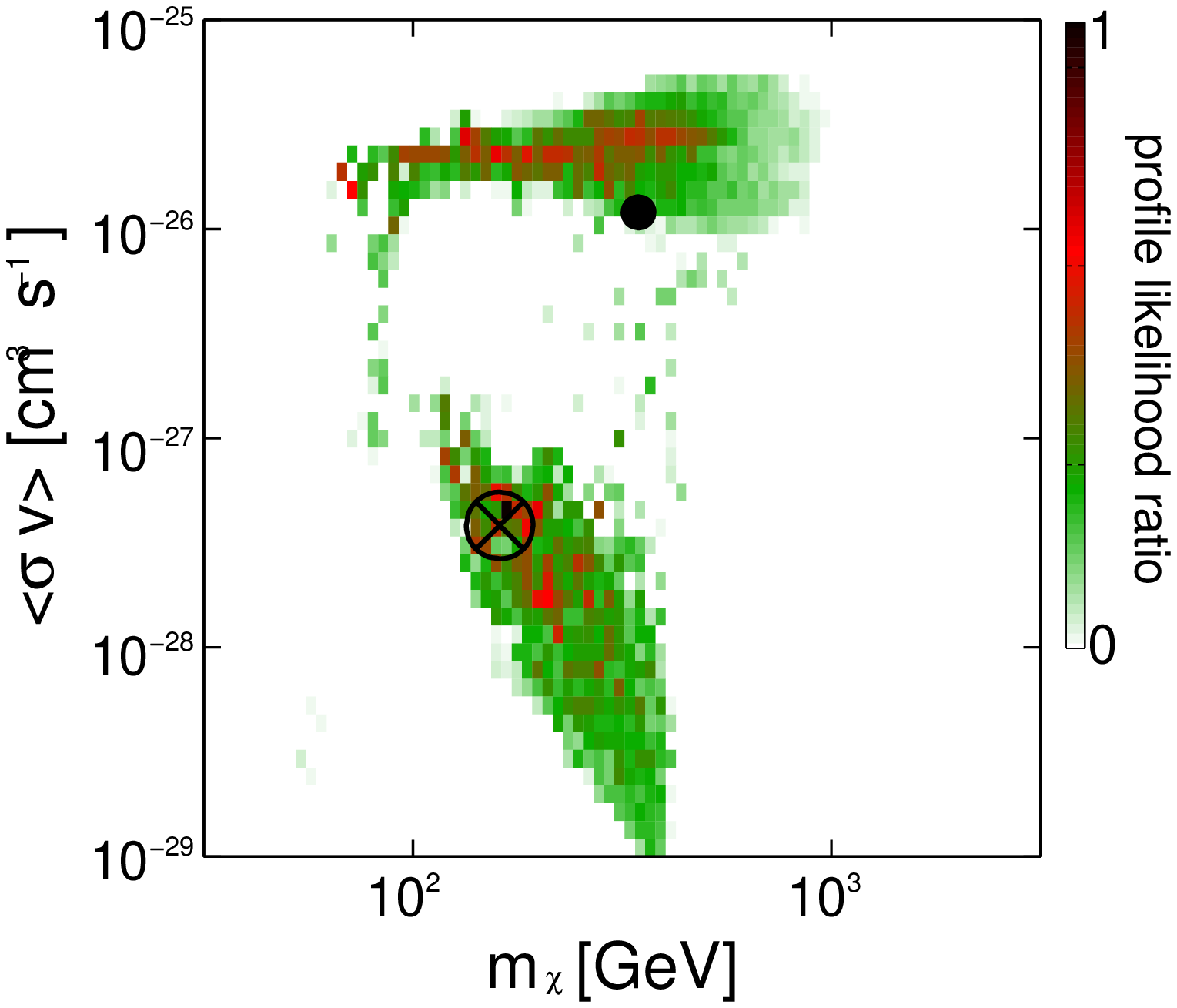}
\includegraphics[width=0.31\linewidth]{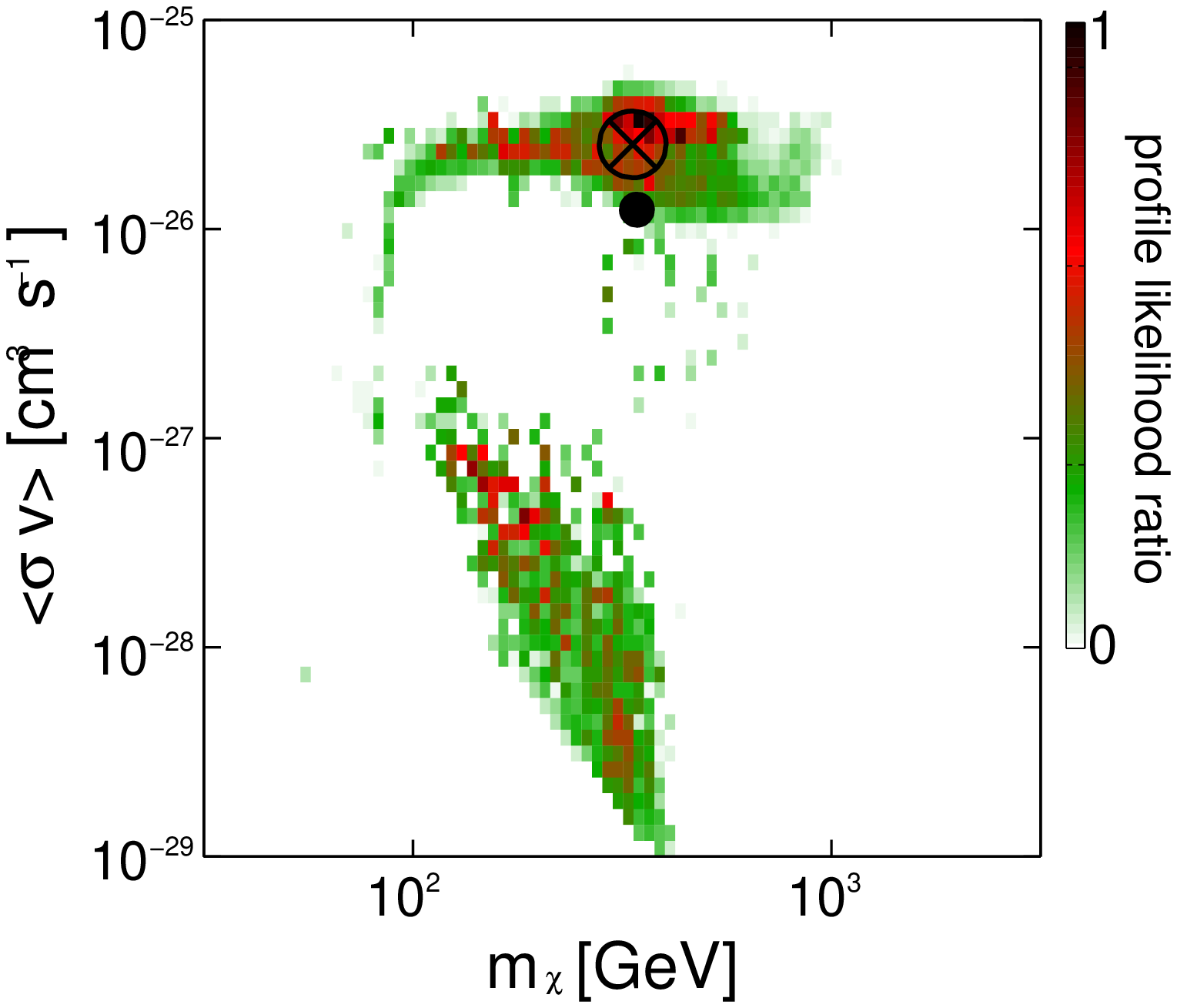}
\includegraphics[width=0.31\linewidth]{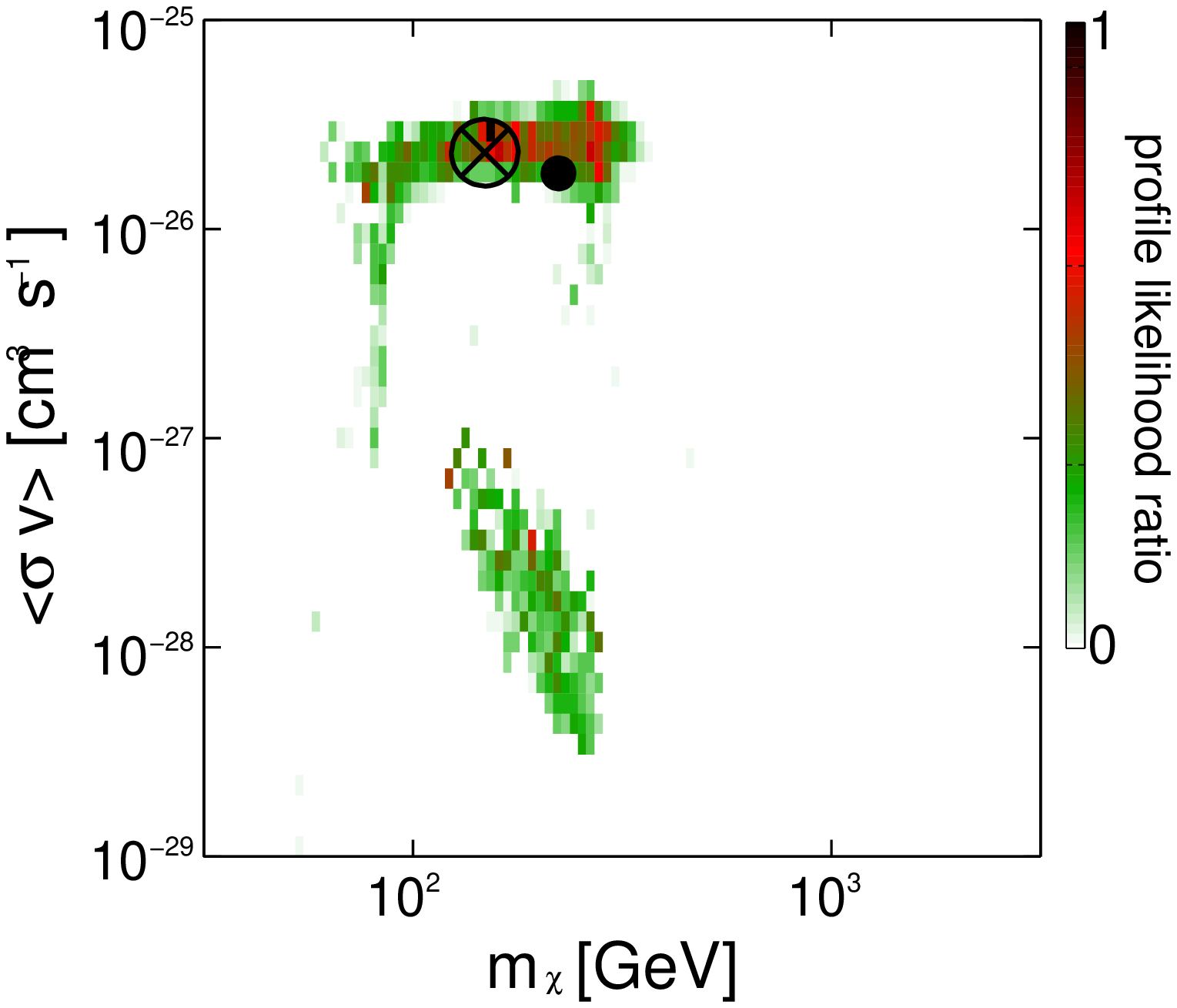}
\includegraphics[width=0.31\linewidth]{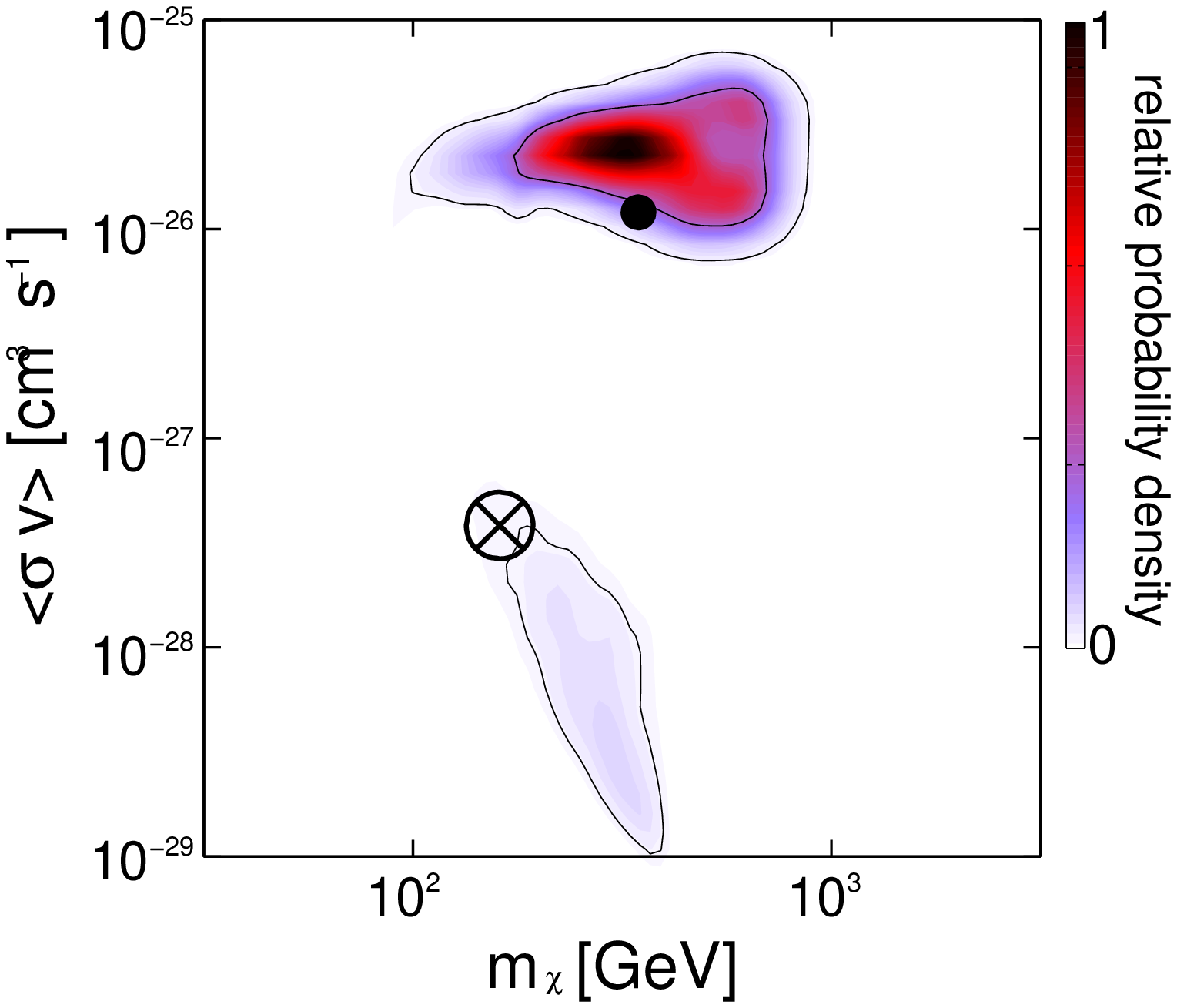}
\includegraphics[width=0.31\linewidth]{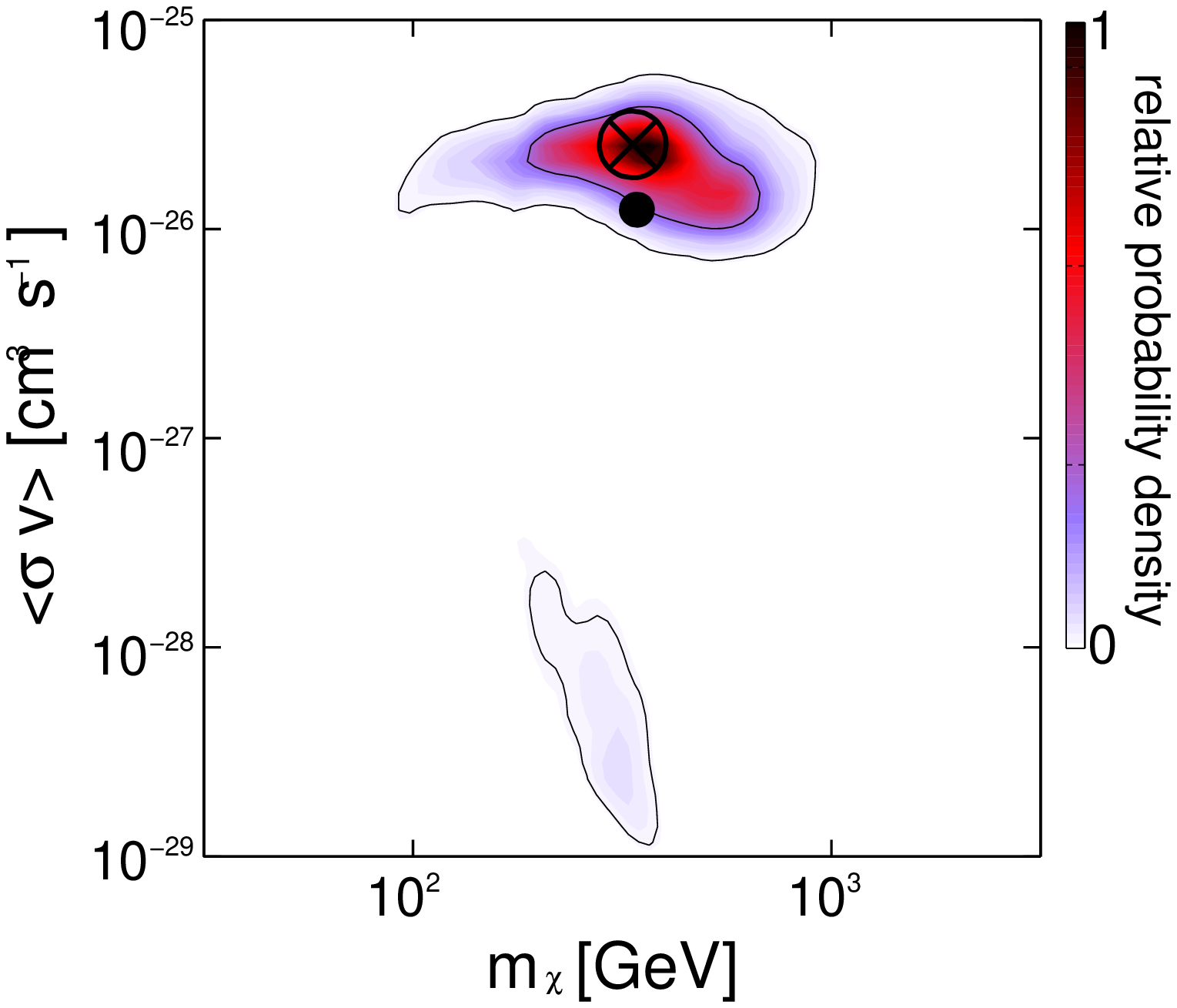}
\includegraphics[width=0.31\linewidth]{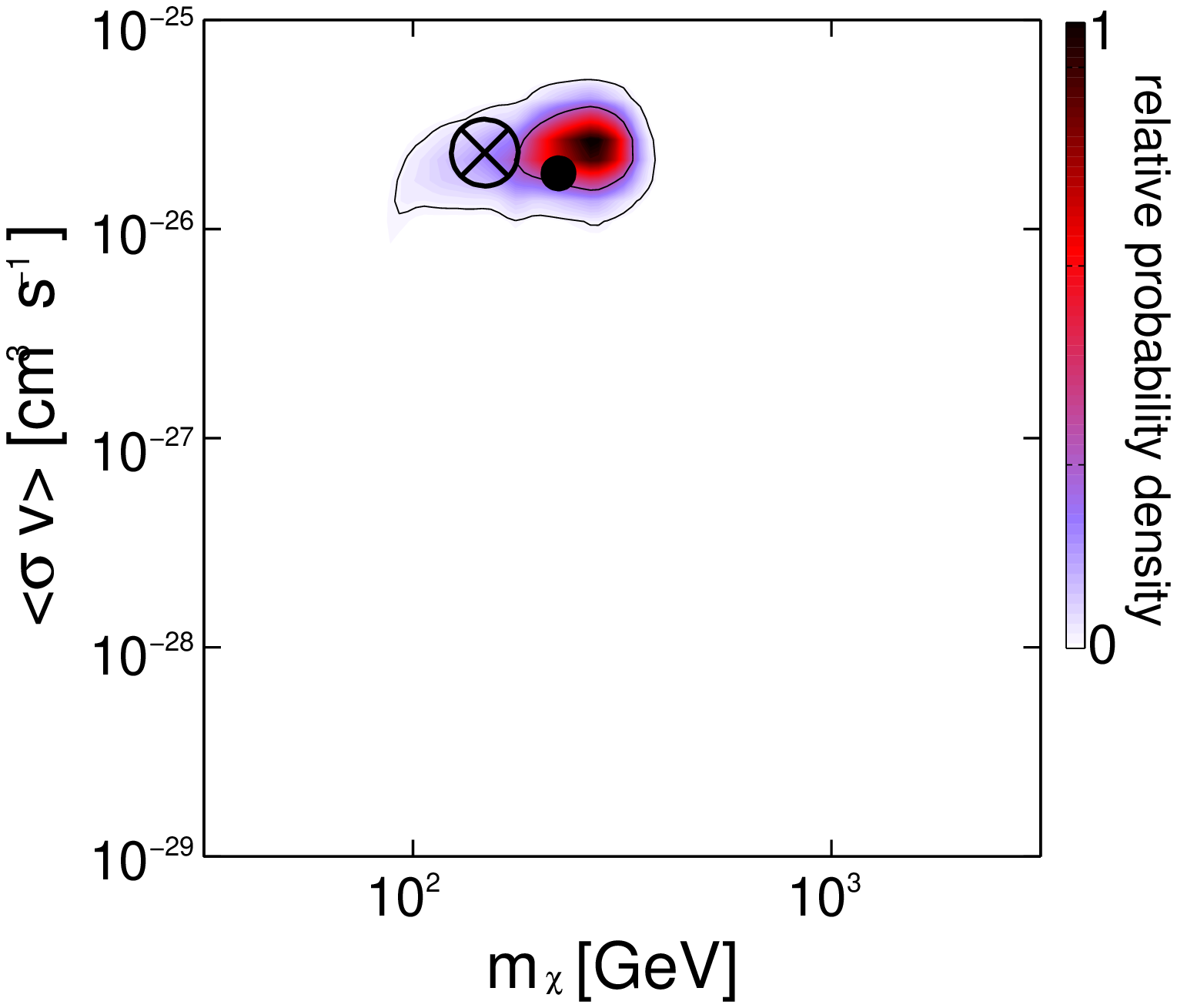}
\end{center} 
\caption{Profile likelihoods $\mathcal{L}/\mathcal{L}_{\text{best
      fit}}$ (upper row) and posterior probability density functions
  (lower row) normalized to the best fit point in the
  $m_{\chi}$-$\langle \sigma v \rangle$ plane. In the left column
  H.E.S.S. data from the Galactic Centre have not been included in the
  scans.  For the other plots, $\bar{J}(\Delta \Omega)\Delta \Omega =
  10 \, \text{sr}$ (middle) and $100 \, \text{sr}$ (right).  The
  $\otimes$ marks the best fit point, while the $\bullet$ marks the
  centre of gravity of the distribution. Contours in the lower plots
  surround $68 \%$ and $95 \%$ credible regions.}
\label{fig_sb2}
\end{figure}

Our analysis is based on SuperBayeS, a package that scans the CMSSM
parameter space and computes various observables, in particular the
gamma-ray spectrum for a given CMSSM model, by interfacing with
DarkSUSY.  It performs statistical inference by comparing the computed
values of the observables to experimental data, using a full
likelihood construction.  SuperBayeS also implements sophisticated
scanning algorithms; in our scans, we chose the MultiNest
\cite{bib_SB4} nested sampling algorithm, with $4000$ live points. In
each iteration step of this algorithm the point with the worst
likelihood in a set of points in parameter space is replaced by a
point with a better likelihood. In order to increase the possibility
to find such a point, the border of the region surrounding all other
points has to be described. This way this region nest the best fit
points iteratively. We used the modified version of SuperBayeS 1.35
described in \cite{bib_Pat} for the analysis of H.E.S.S. data,
supplemented with an appropriate H.E.S.S. likelihood term (given
below).  This modified version employs DarkSUSY 5.0.4 for the
calculation of relic densities and gamma-ray spectra, including the
full calculation of internal bremsstrahlung (both VIB and FSR)
\cite{bib_IB} crucial for our analysis.

We scanned over $60 \, \text{GeV} \leq m_0 \leq 4000 \, \text{GeV}$,
$60 \, \text{GeV} \leq m_\frac12 \leq 4000 \, \text{GeV}$, $-7000 \leq
A_0 \leq 7000$ and $2 \leq \tan\beta \leq 65$, setting $\mu>0$ and
applying linear priors to the other parameters. We also scanned over
the top and bottom quark masses, and the strong and electromagnetic
coupling constants, treating them as SM nuisance parameters.  We
incorporate the effects of the nuisance parameters in our analysis by
either computing the profile likelihood (i.e.  maximising the
likelihood with respect to the nuisance parameters, at each point in
the CMSSM parameter space), or by marginalising over them, integrating
the posterior distribution at each point in the CMSSM parameter space
over the nuisance space.  Similarly, we choose to present
distributions and likelihoods for subsets of CMSSM parameters by
further profiling or marginalising over the remaining CMSSM
parameters.

We use the unfolded (deconvolved; see \cite{bib_hessGC5}) H.E.S.S.
spectrum to directly compare with the theoretically predicted
gamma-ray spectrum.  We also use the observables, experimental
likelihoods and SM nuisance likelihoods described in
Ref.~\cite{bib_SB6}; these are also the same as we employed in
Refs.~\cite{bib_Yashar, bib_Pat}.  In particular, we compared the
relic density to data from the 5-year Wilkinson Microwave Anisotropy
Probe (WMAP), which found $\Omega_{\text{DM}} h^{2} = 0.1099 \pm
0.0062$ at the $1\sigma$ level \cite{bib_WMAP5}.  Other observables
were: LEP constraints on sparticle masses and the Higgs mass,
measurements of the anomalous magnetic moment of the muon $(g - 2)$,
the mass difference $m_{\bar{B}_{s}} - m_{B_{s}}$, and the branching
fractions of the rare processes $b \rightarrow s \gamma$, $\bar{B}_{u}
\rightarrow \nu \tau^{-}$ and $\bar{B}_{s} \rightarrow \mu^{+}
\mu^{-}$.

The likelihood of one CMSSM model is defined by
\begin{equation}
- \ln \mathcal{L} = \sum_{i} - \ln \mathcal{L}_{i}
\end{equation}
where $\mathcal{L}_{i}$ is the likelihood associated with each
individual observable. For the H.E.S.S. spectrum from the GC, we used
\begin{equation}
- \ln \mathcal{L}_{\text{H.E.S.S., GC}} = \frac{\chi^{2}}{2}
\end{equation}
with the $\chi^{2}$ described in the previous subsection.

In Figures \ref{fig_sb1} and \ref{fig_sb2} we show both the profile
likelihood and the posterior probability density function (assuming
flat priors) for three different scans. The value of $\bar{J}(\Delta
\Omega) \Delta \Omega$ increases left to right from $0 \, \text{sr}$
-- no dark matter in the GC region, no H.E.S.S. data included in the
scan -- to $100 \, \text{sr}$ in the last column. Figure \ref{fig_sb1}
shows the results of the scans projected down into the
$m_\frac12$-$m_{0}$ plane, while Figure \ref{fig_sb2} shows the
resulting distributions in the $m_{\chi}$-$\langle \sigma v \rangle$
plane.

We see from Figures \ref{fig_sb1} and \ref{fig_sb2} that with
increasing $\bar{J}(\Delta \Omega)\Delta \Omega$, the likelihoods of
points in the coannihilation region and the higher-mass part of the
focus point are reduced.  This can also be seen in the movement of the
best-fit point from the tip of the coannihilation region to a low-mass
part of the focus point when GC data are introduced.  In contrast, the
posterior mean does not move substantially when GC data are included
in fits, reflecting the fact that the focus point carries the majority
of the posterior mass when linear priors are employed, and is left
largely intact after the application of GC data.

However, the values we have used for $\bar{J}(\Delta \Omega)\Delta
\Omega$ in these scans with GC data are unrealistically large. The GC
source delivers a strong astrophysical background, hindering dark
matter investigations. The scanning technique will be useful for
future observations of the GC region however, especially with upcoming
experiments like the \v{C}erenkov Telescope Array (CTA).

\section{CMSSM likelihood scan with data from the Sagittarius dwarf
  galaxy} \label{subsec_SgD}

\begin{figure}[ht]  
\begin{center}
\includegraphics[width=0.31\linewidth]{noHess_2D_profl_1.eps}
\includegraphics[width=0.31\linewidth]{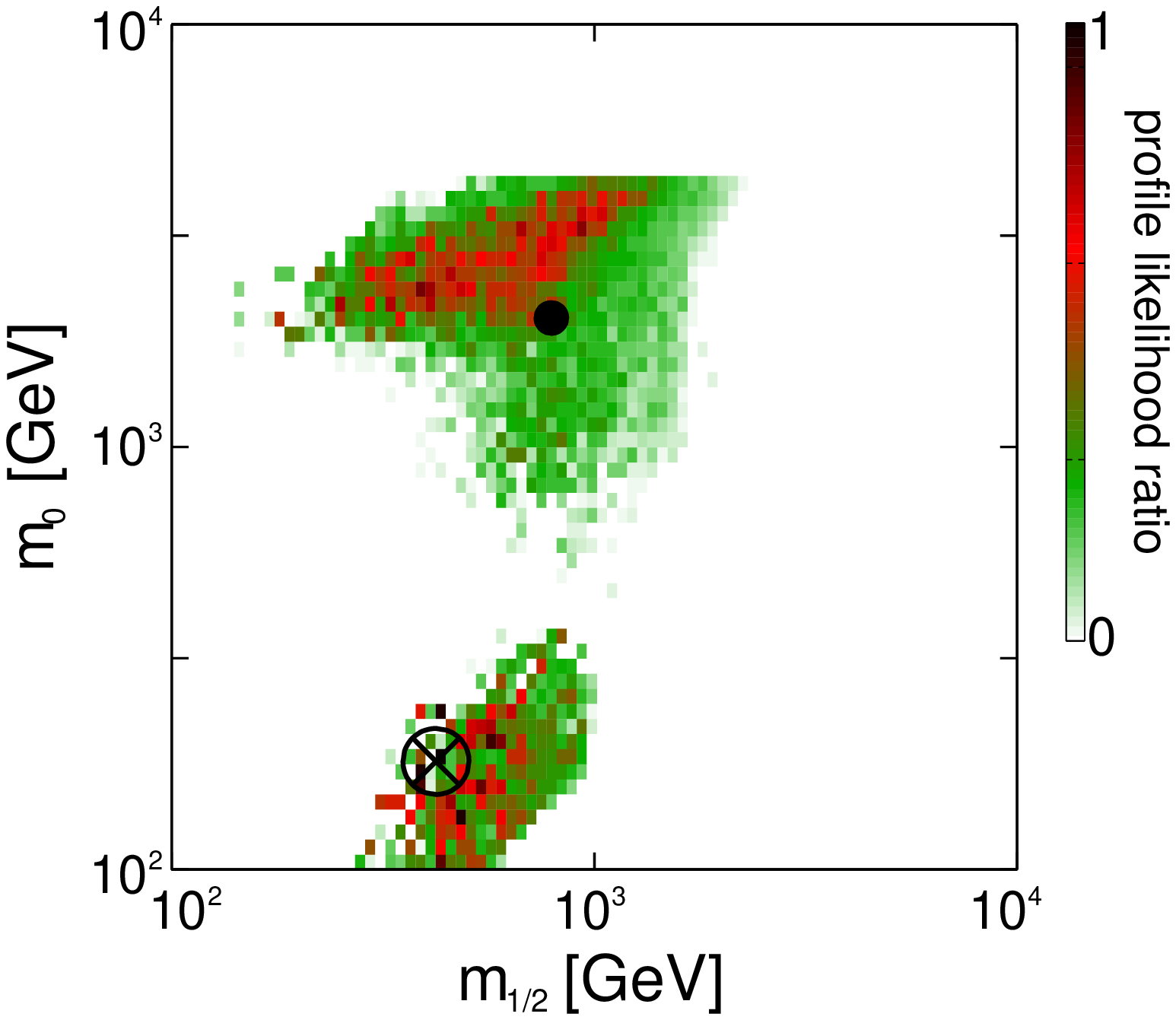}
\includegraphics[width=0.31\linewidth]{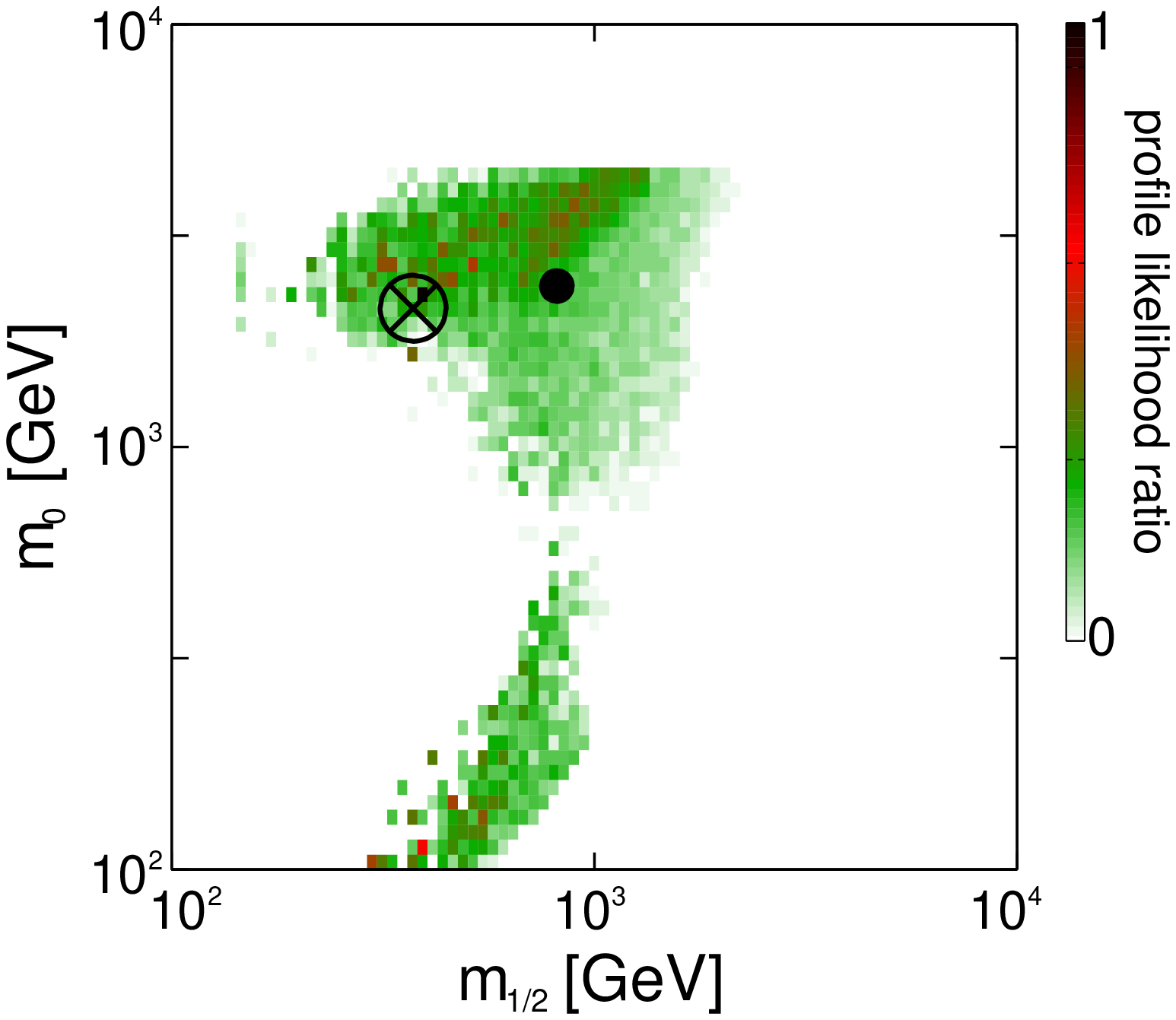}
\includegraphics[width=0.31\linewidth]{noHess_2D_marg_1.eps}
\includegraphics[width=0.31\linewidth]{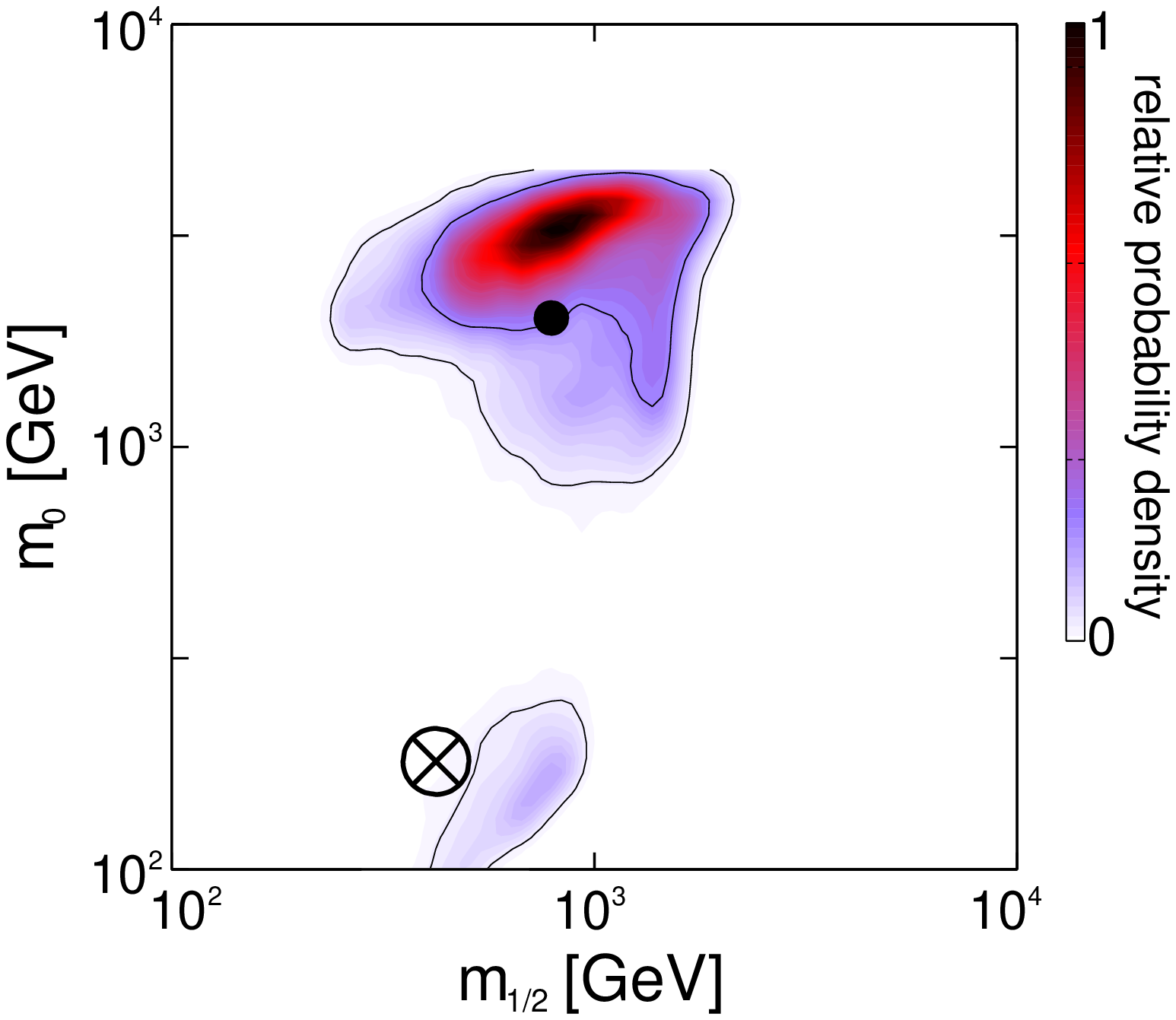}
\includegraphics[width=0.31\linewidth]{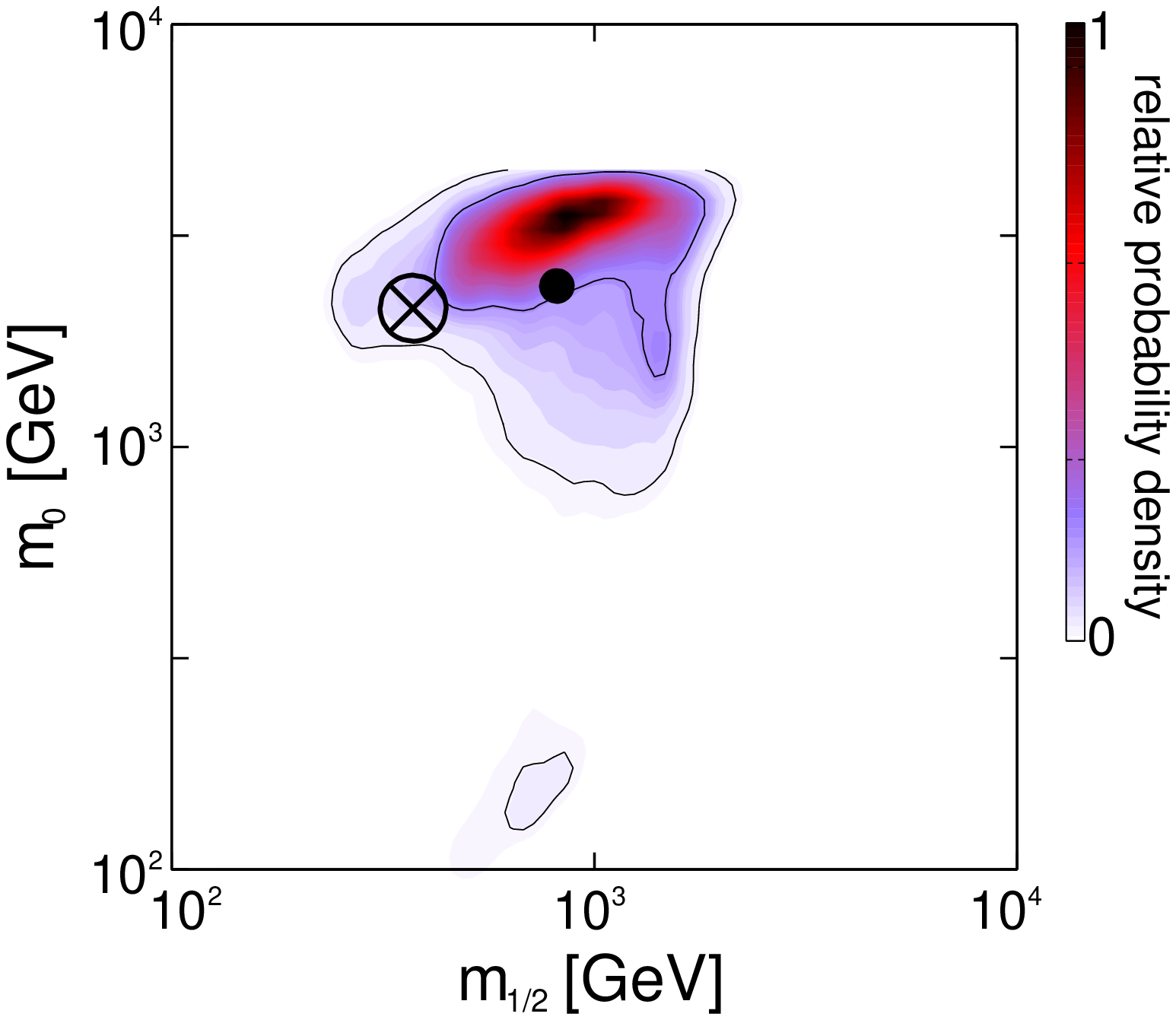}
\end{center} 
\caption{Profile likelihood ratios
  $\mathcal{L}/\mathcal{L}_{\text{best fit}}$ (upper row) and
  posterior probability density functions (lower row) normalized to
  the best fit point in the $m_\frac12$-$m_{0}$ plane. In the left
  column H.E.S.S. data from the (SgrD) have not been included in the
  scans.  For the other plots, we assume an NFW (middle) or a cored DM
  profile (right).  The $\otimes$ marks the best fit point, while the
  $\bullet$ marks the centre of gravity of the distribution.  The
  contours in the lower row plots surround the $68 \%$ and the $95 \%$
  CL regions.}
\label{fig_sgd1}
\end{figure}

\begin{figure}[ht]  
\begin{center}
\includegraphics[width=0.31\linewidth]{noHess_2D_profl_7.eps}
\includegraphics[width=0.31\linewidth]{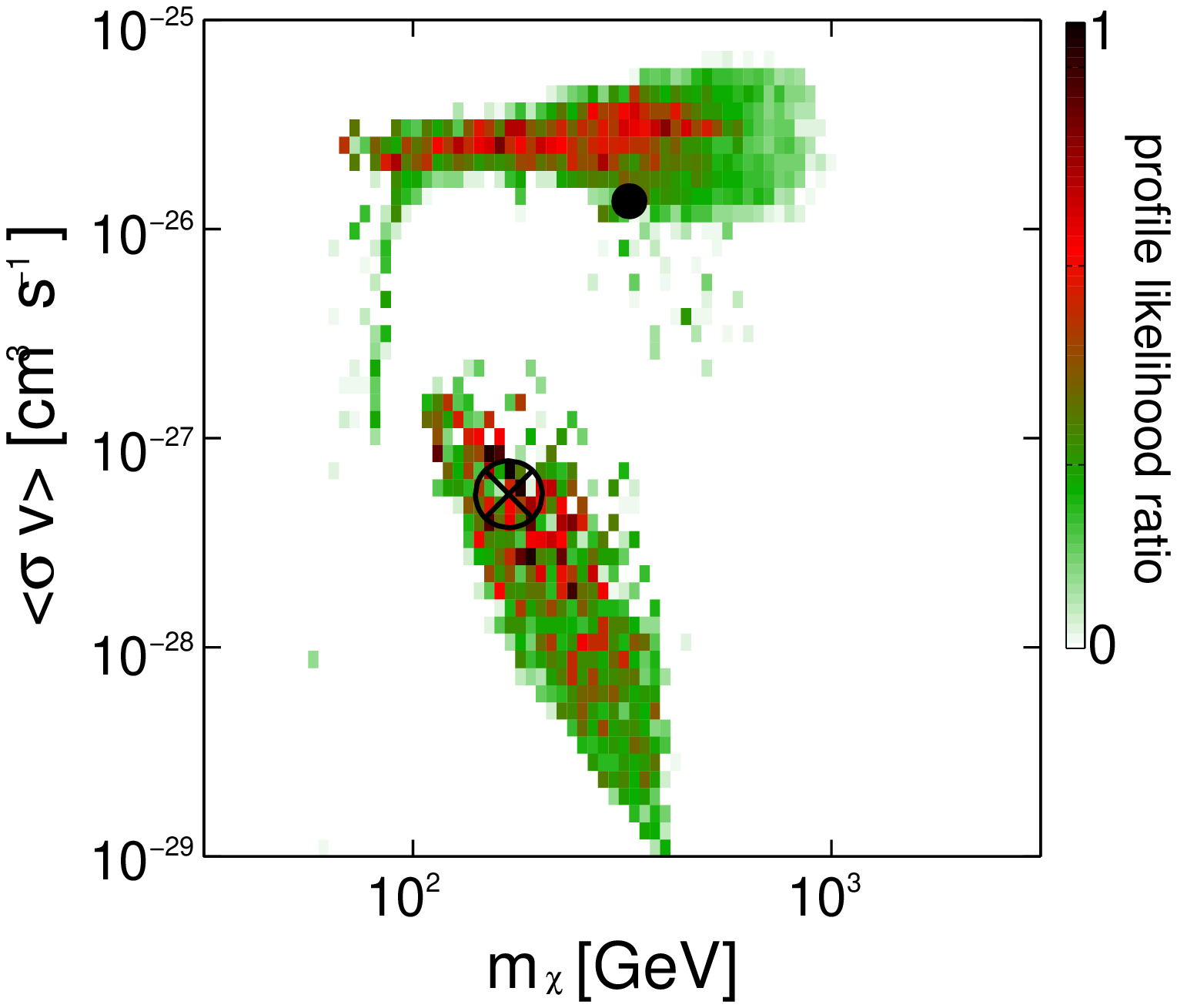}
\includegraphics[width=0.31\linewidth]{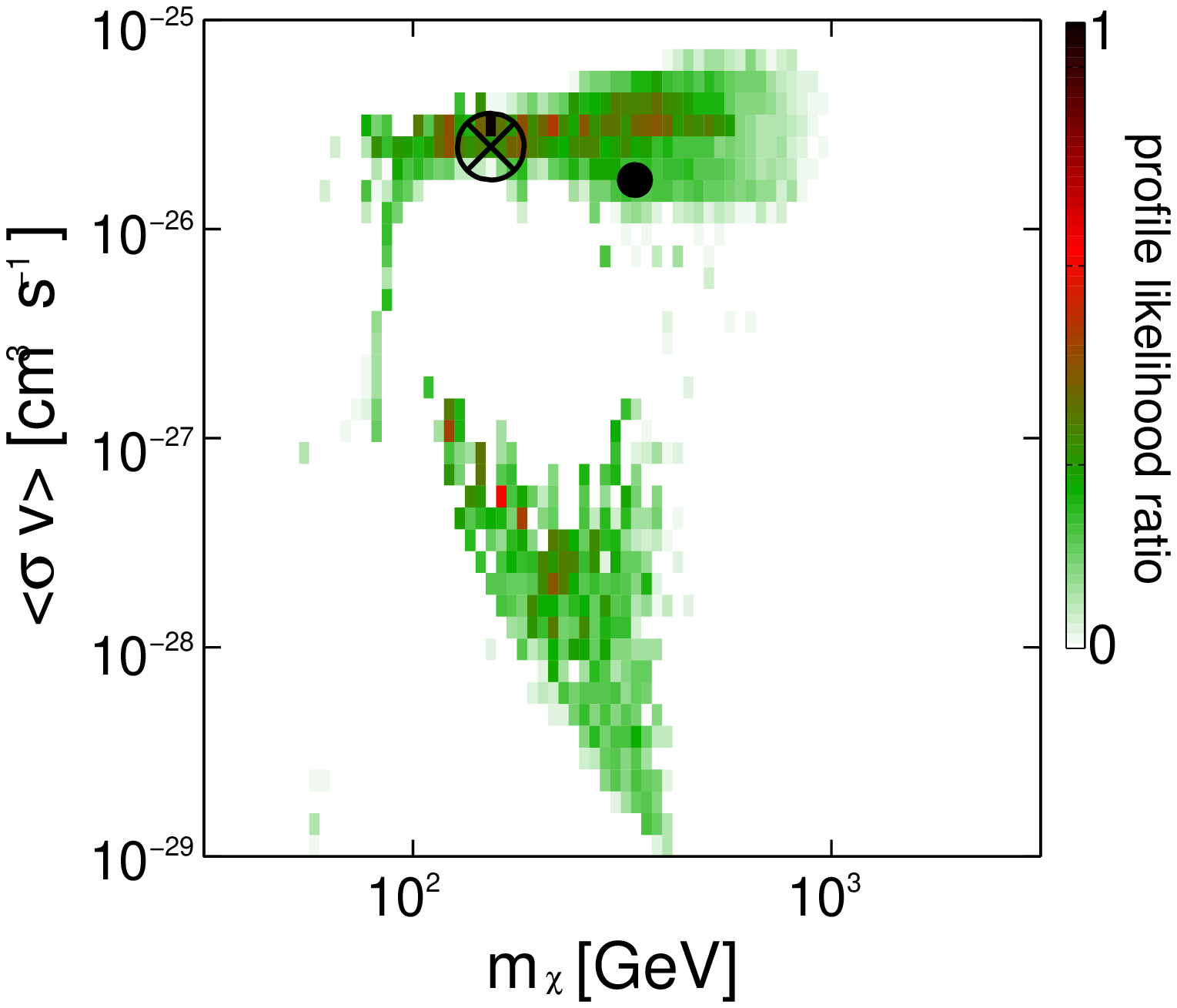}
\includegraphics[width=0.31\linewidth]{noHess_2D_marg_7.eps}
\includegraphics[width=0.31\linewidth]{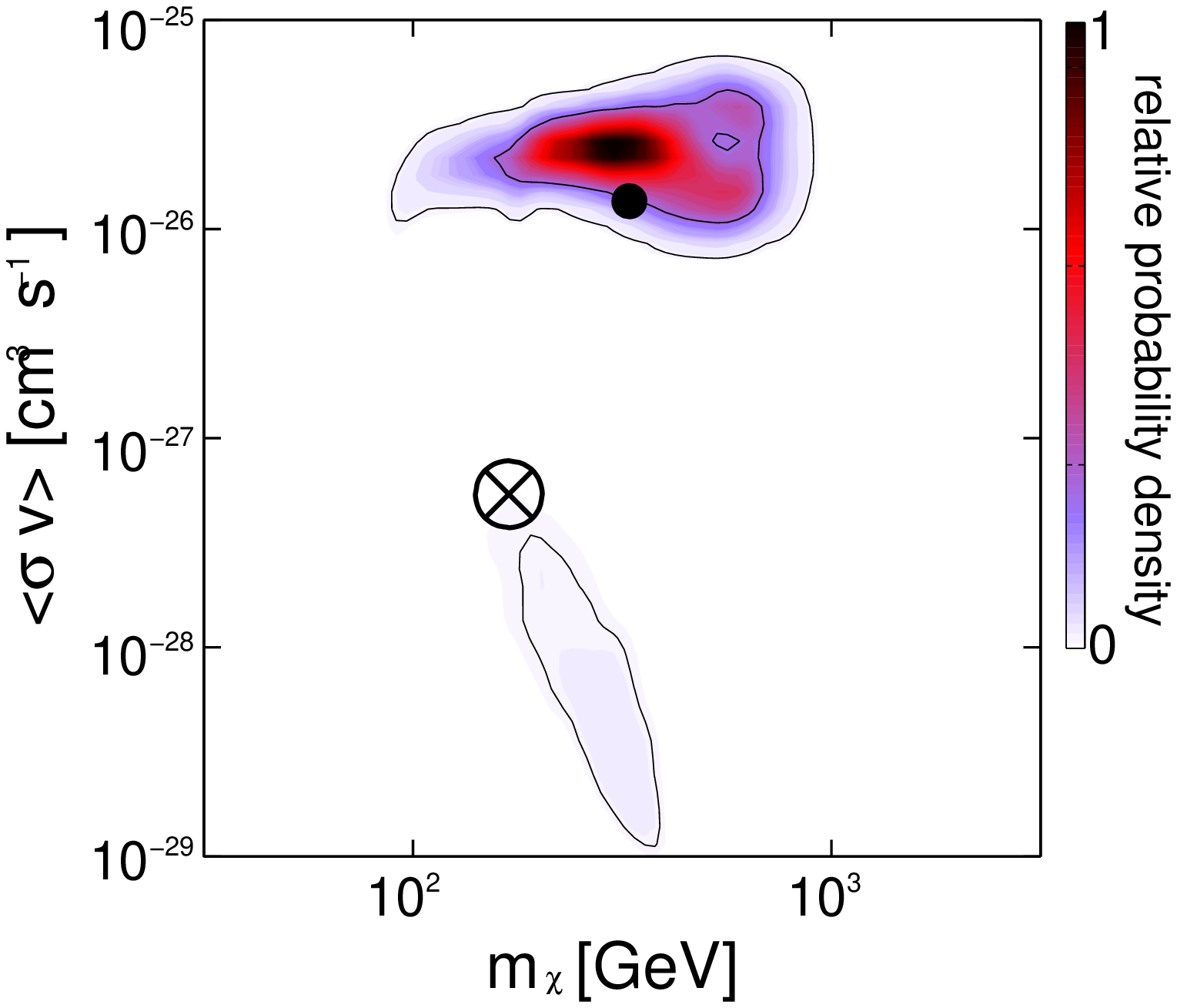}
\includegraphics[width=0.31\linewidth]{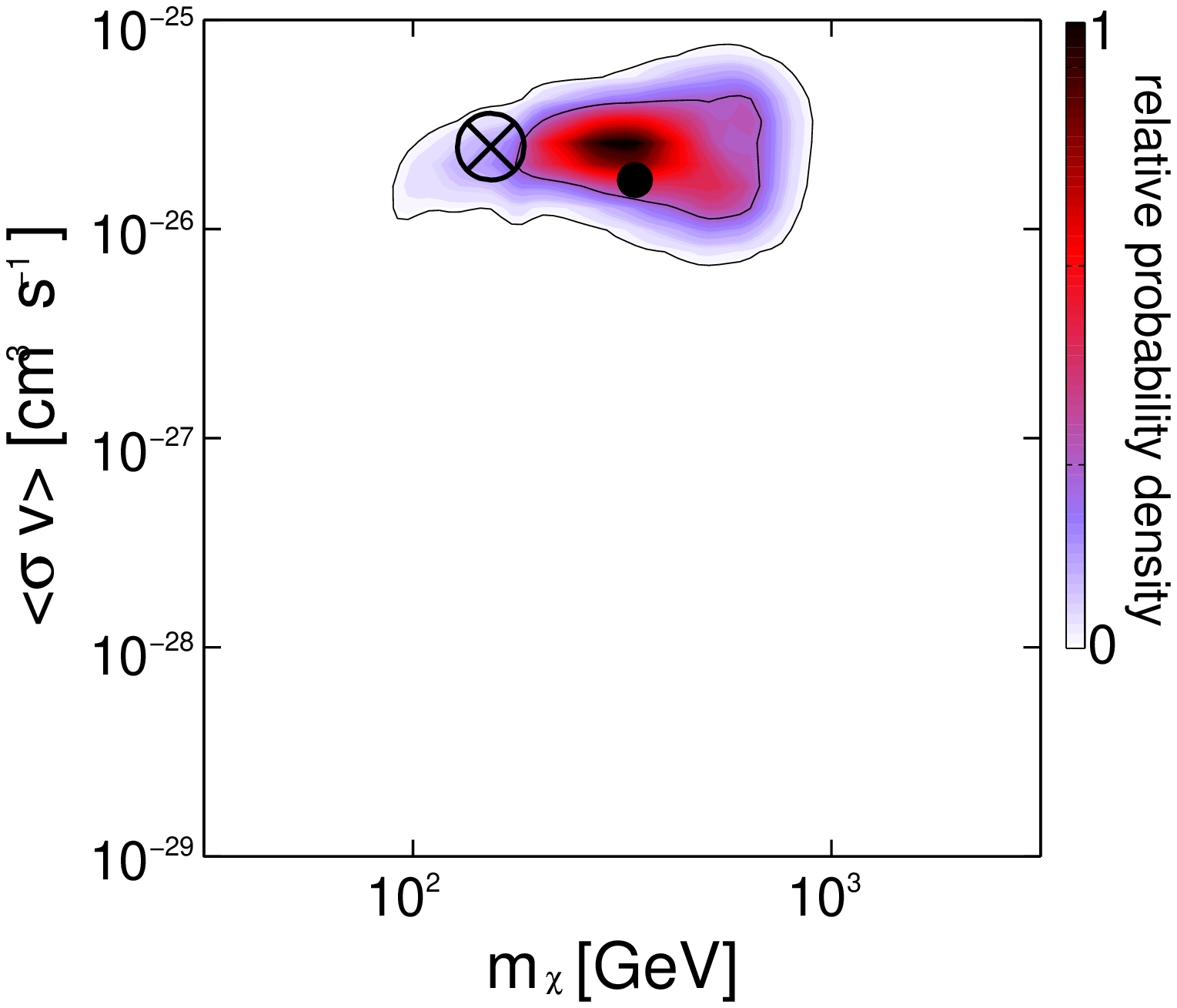}
\end{center}
\caption{Profile likelihood ratios
  $\mathcal{L}/\mathcal{L}_{\text{best fit}}$ (upper row) and
  posterior probability density functions (lower row) normalized to
  the best fit point in the $m_{\chi}$-$\langle \sigma v \rangle$
  plane. In the left column H.E.S.S. data from the SgrD have not been
  included in the scans.  For the other plots, we assume an NFW
  (middle) or a cored DM profile (right). The $\otimes$ marks the best
  fit point, while the $\bullet$ marks the center of gravity of the
  distribution.  The contours in the lower row plots surround the $68
  \%$ and the $95 \%$ CL regions.}
\label{fig_sgd2}
\end{figure}

We assume two different DM profiles for the SgrD: a (cuspy) NFW and a
cored profile. The first one delivers a scale factor of
$\bar{J}(\Delta \Omega)\Delta \Omega \lvert_{\text{NFW}} =
0.0186$\,\text{sr}, whereas the second one gives $\bar{J}(\Delta
\Omega)\Delta \Omega \lvert_{\text{cored}} = 0.636$\,\text{sr} (with
the definition in equation \ref{eq_dmflux}). Although it is less
concentrated very close to the centre, the cored halo gives a larger
$J$ factor because its dark matter core radius is only 1.5\,pc, which
is within the observed solid angle. The small core radius leads to a
steeper profile than NFW beyond $r=1.5\,pc$, and a higher DM density
around $r\sim1.5\,pc$.  The calculated upper limits just begin to
touch interesting parts of parameter space (see the Erratum to
\cite{bib_hessSGD}).

Because SgrD experiences heavy tidal disruptions, there are large
uncertainties in its density profile, leading to a large uncertainty
in the resultant $J$ factor.  We refer the reader to e.g.
Refs.~\cite{bib_hessSGD,bib_Viana} and references therein for more
extensive discussions.  Because of this uncertainty, in the following
sections we also investigate the impacts of H.E.S.S. observations of
the Galactic diffuse emission and other dwarf galaxies on the CMSSM,
for which the $J$ factors are better constrained.

In order to subtract the large cosmic ray background, this is
estimated via a dedicated OFF-region in the H.E.S.S. analysis. This
region might also contain a significant fraction of a hypothetical DM
annihilation signal, which in this case would also be subtracted
\cite{bib_Mack}.  Since more than $90 \%$ of the DM signal of both the
density profiles that we consider here originates from inside the
ON-region (see \cite{bib_hessSGD}), this effect is negligible in our
case.

To include these data into our likelihood calculation, we need an
estimate of the flux and its error. $437$ events were observed by
H.E.S.S.  in the ``ON-region'' centred on SgrD, and in the surrounding
annular ``OFF-region'' $4270$ events were collected. Since there is a
difference of a factor of $10.1$ in the areas of the sky covered by
these two regions, there are $14.2$ excess events observed in the
ON-region. This is not statistically significant. Assuming that
observed events follow Poisson statistics, the actual observed flux
and its error are $\Phi(E>250 \, \text{GeV}) = (0.9 \pm 1.4) \cdot
10^{-12} \, \text{cm}^{-2} \text{s}^{-1}.$ We use this for the
calculation of a (gaussian) likelihood.

Calculating the expected integrated flux from each CMSSM model and
comparing with this value delivers us an easy estimate of the
likelihood
\begin{equation}
  - \ln \mathcal{L}_{\text{H.E.S.S., Sag}} = 
  \frac{(\Phi_{\text{measured}}-\Phi_{\text{model}})^{2}}{2 \sigma_{\Phi}^{2}}
\end{equation}

The results of these scans (with somewhat more realistic density
profiles than we employed for the GC) can be seen in Figures
\ref{fig_sgd1} and \ref{fig_sgd2}. We see that the coannihilation
region becomes steadily more disfavoured for increasing $J$, due to
the large virtual IB signal produced by models in this region.  In
general the only observable that strongly favours the coannihilation
region over higher sparticle masses (as found in e.g.  the focus point
region) is $g-2$, the anomalous magnetic moment of the muon
\cite{bib_SB6}.  When H.E.S.S. observations of the SgrD are included
in the total likelihood, we see that their preference for the focus
point over the stau coannihilation region essentially nullifies the
impact of $g-2$.  This allows $b\rightarrow s\gamma$ to more clearly
exert its preference for higher sparticle masses, leading to a
stronger preference for focus point SUSY over the stau coannihilation
region.

\section{CMSSM likelihood scan with two other dwarf spheroidal galaxies}
\label{sec_2dwarfs}

\begin{figure}[ht]
\begin{center}
\includegraphics[width=0.31\linewidth]{noHess_2D_profl_1.eps}
\includegraphics[width=0.31\linewidth]{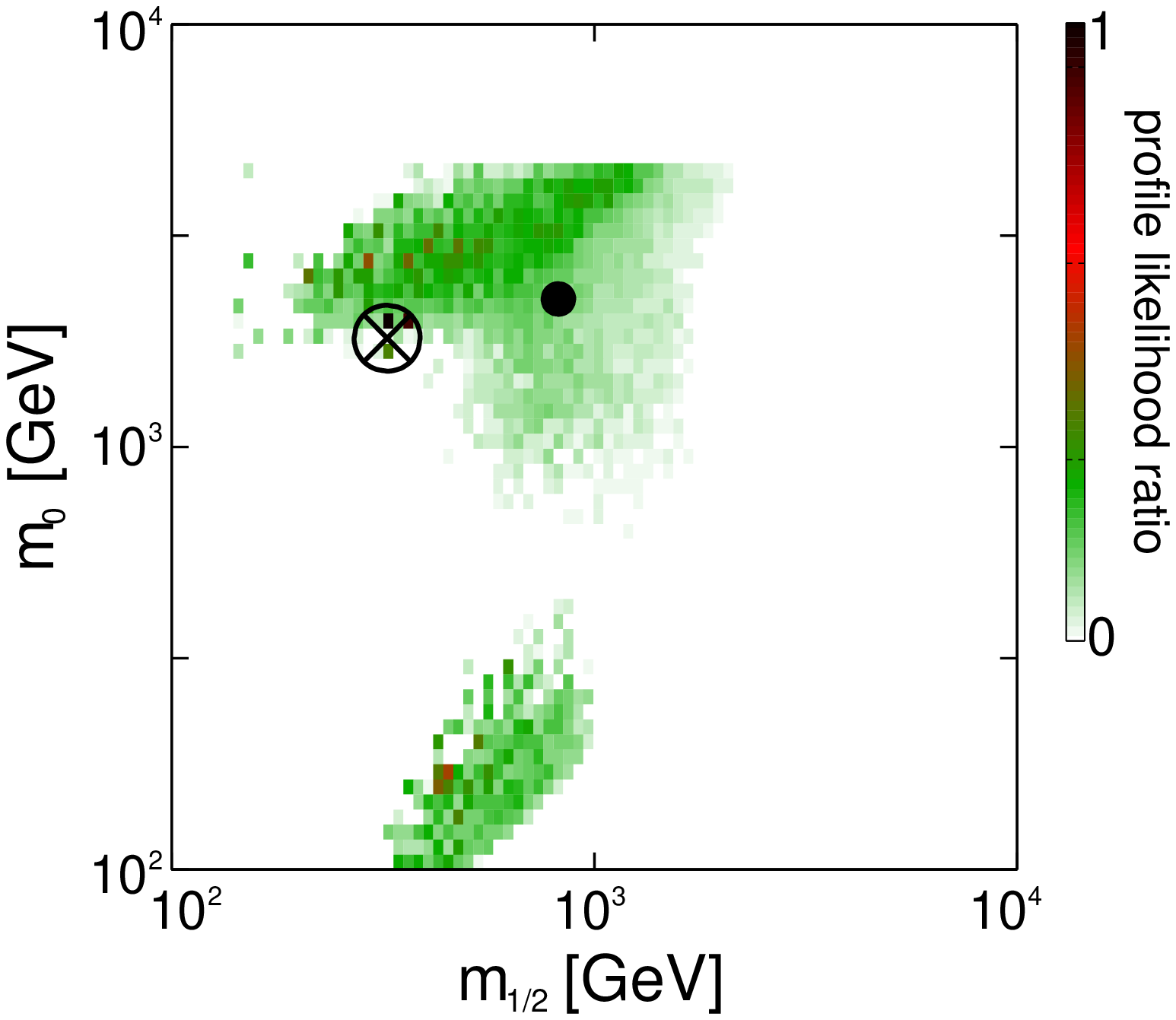}
\includegraphics[width=0.31\linewidth]{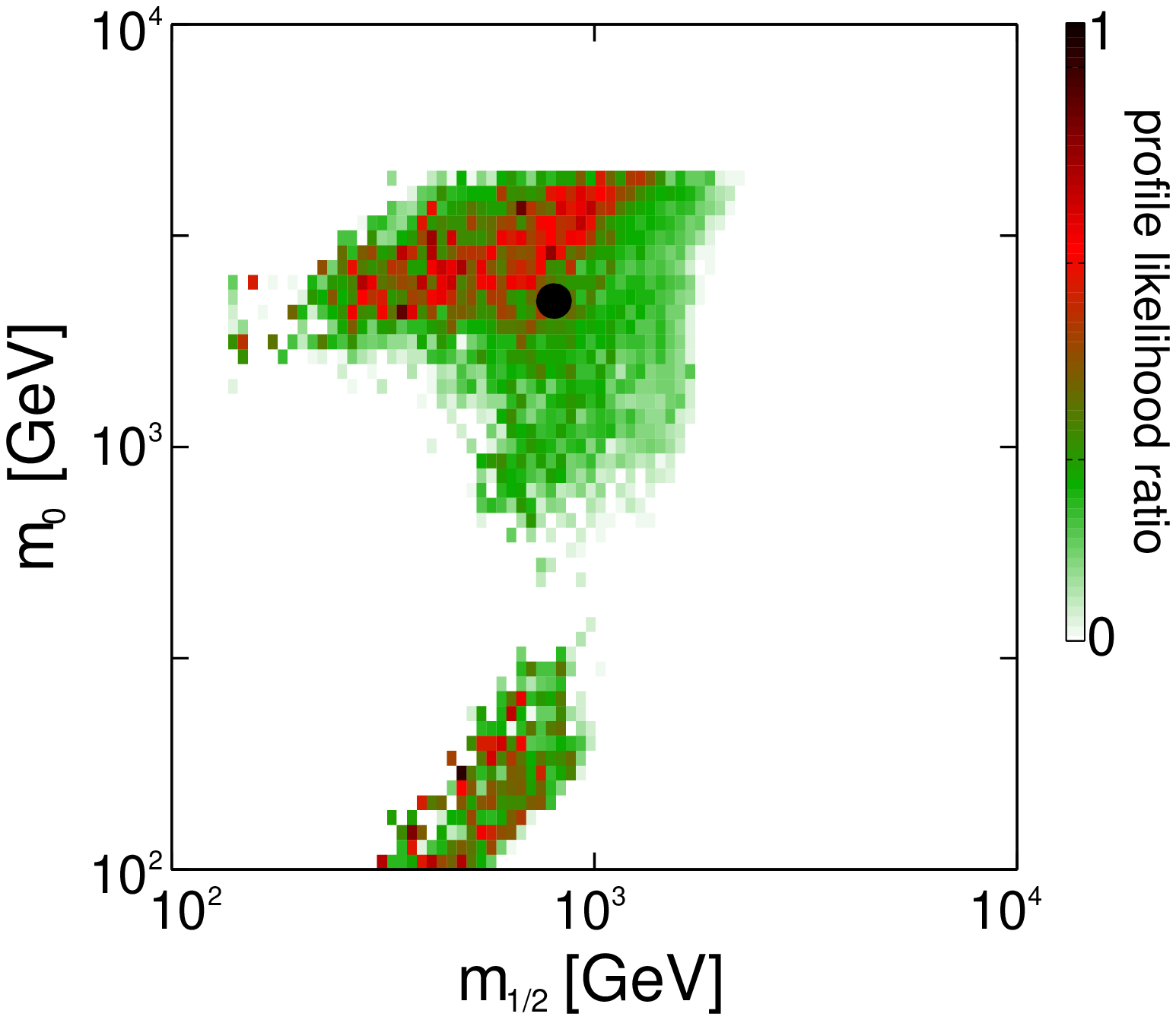}
\includegraphics[width=0.31\linewidth]{noHess_2D_marg_1.eps}
\includegraphics[width=0.31\linewidth]{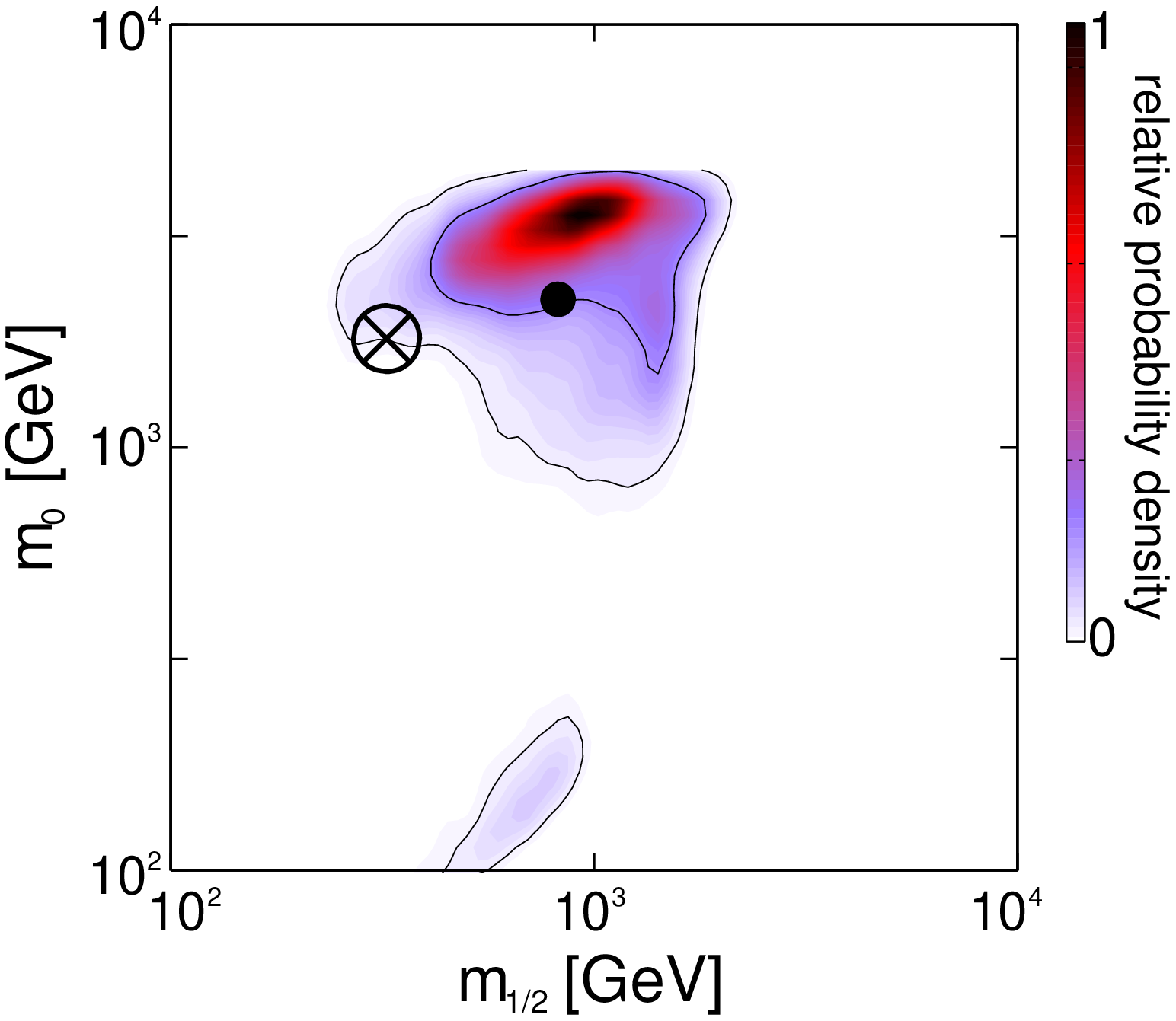}
\includegraphics[width=0.31\linewidth]{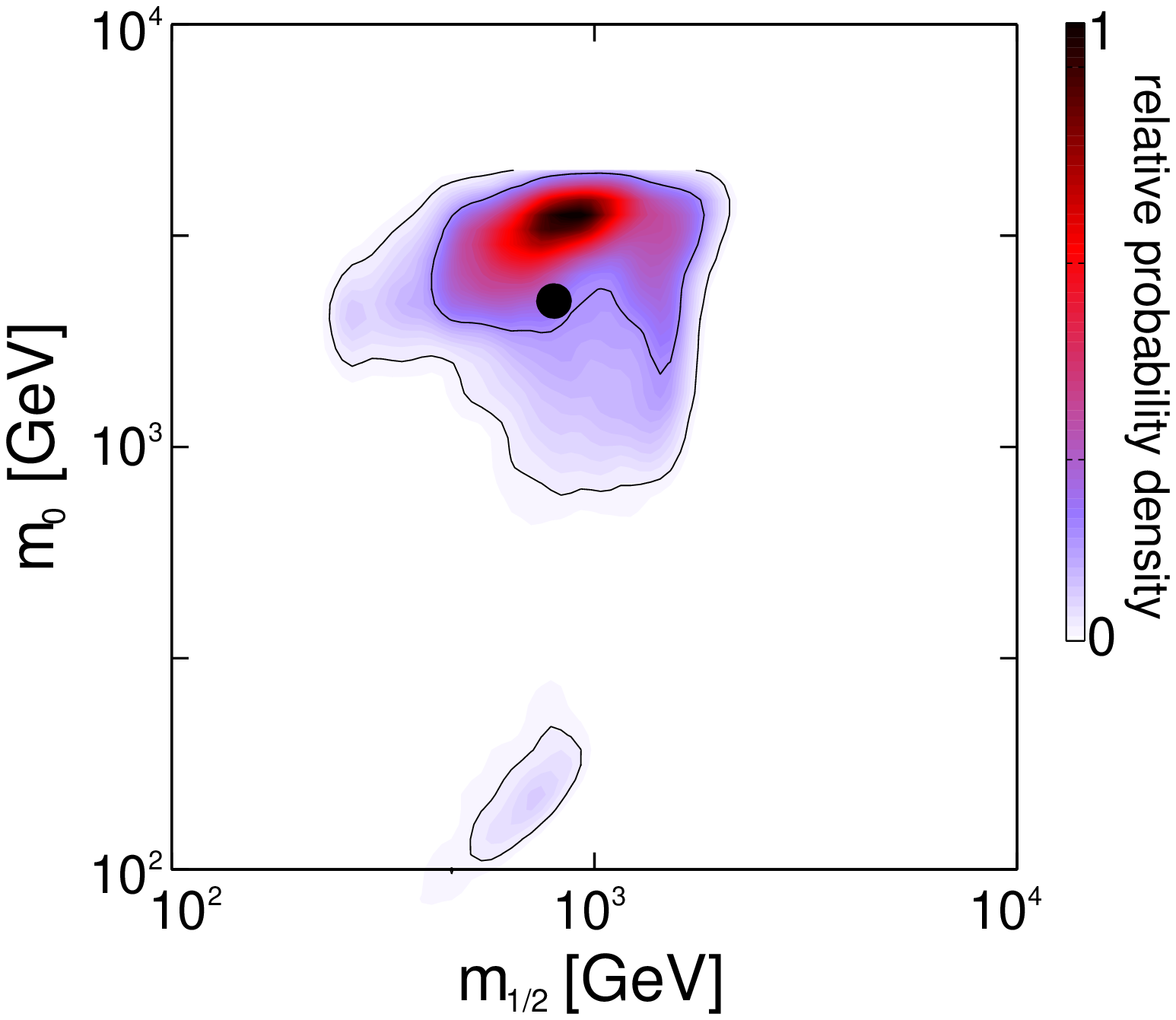}
\end{center}
\caption{Profile likelihood ratios
  $\mathcal{L}/\mathcal{L}_{\text{best fit}}$ (upper row) and
  posterior probability density functions (lower row) normalized to
  the best fit point in the $m_\frac12$-$m_{0}$ plane. In the left
  column no H.E.S.S. data are included in the scans. In the middle
  column Carina and Sculptor data are included.  In the right column
  SgrD is also included.}
\label{fig_dwarfs1}
\end{figure}

\begin{figure}[ht]
\begin{center}
\includegraphics[width=0.31\linewidth]{noHess_2D_profl_7.eps}
\includegraphics[width=0.31\linewidth]{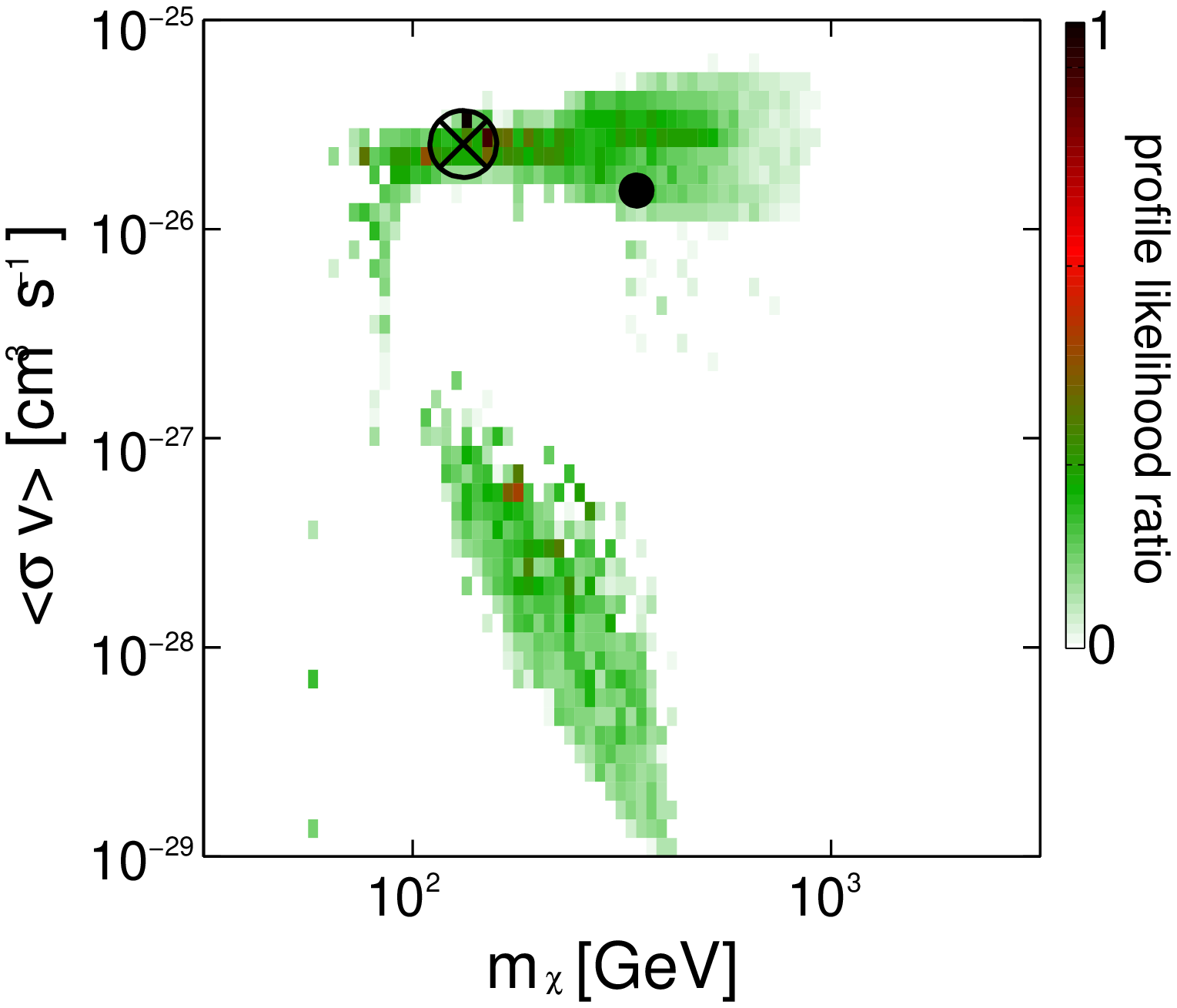}
\includegraphics[width=0.31\linewidth]{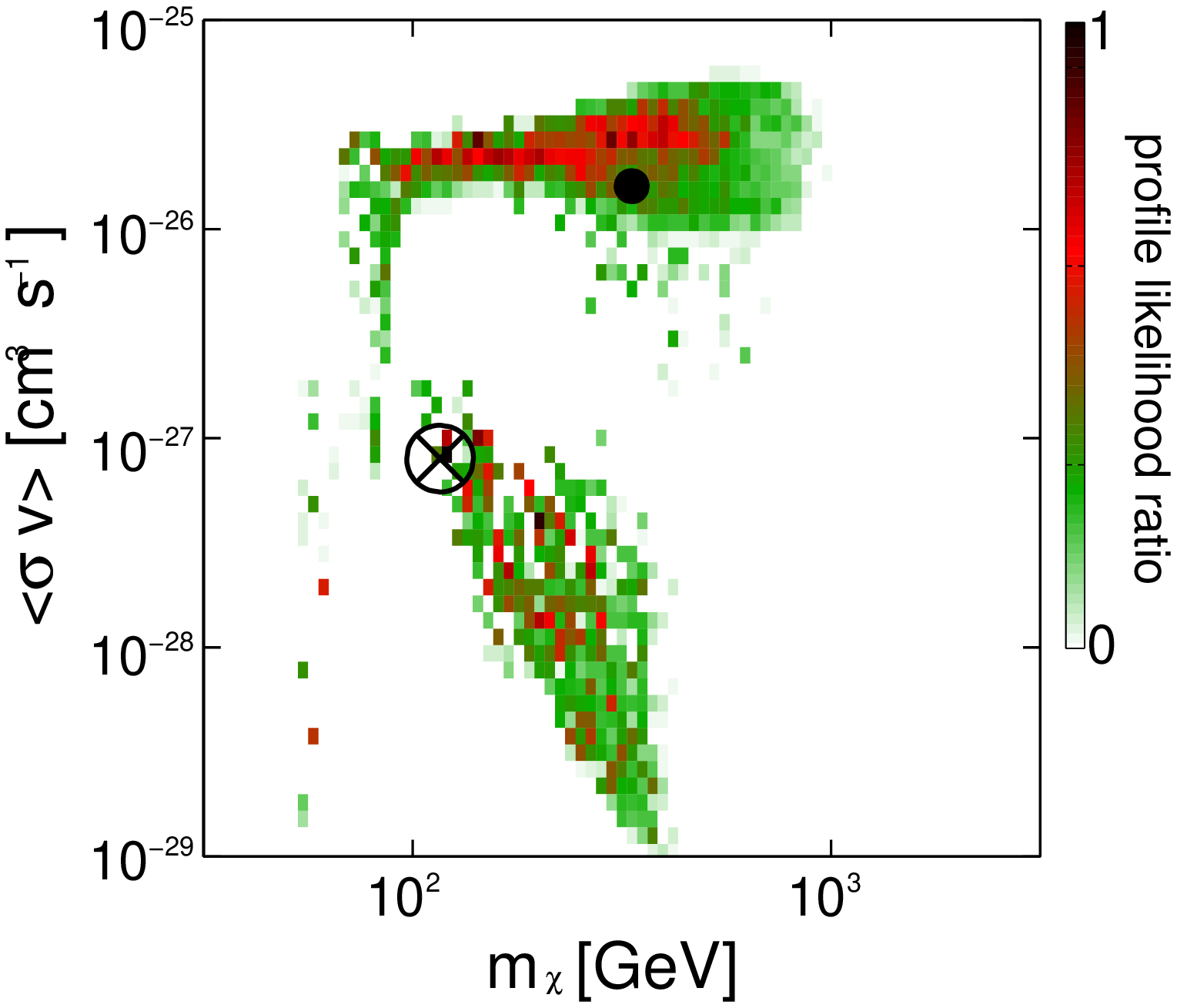}
\includegraphics[width=0.31\linewidth]{noHess_2D_marg_7.eps}
\includegraphics[width=0.31\linewidth]{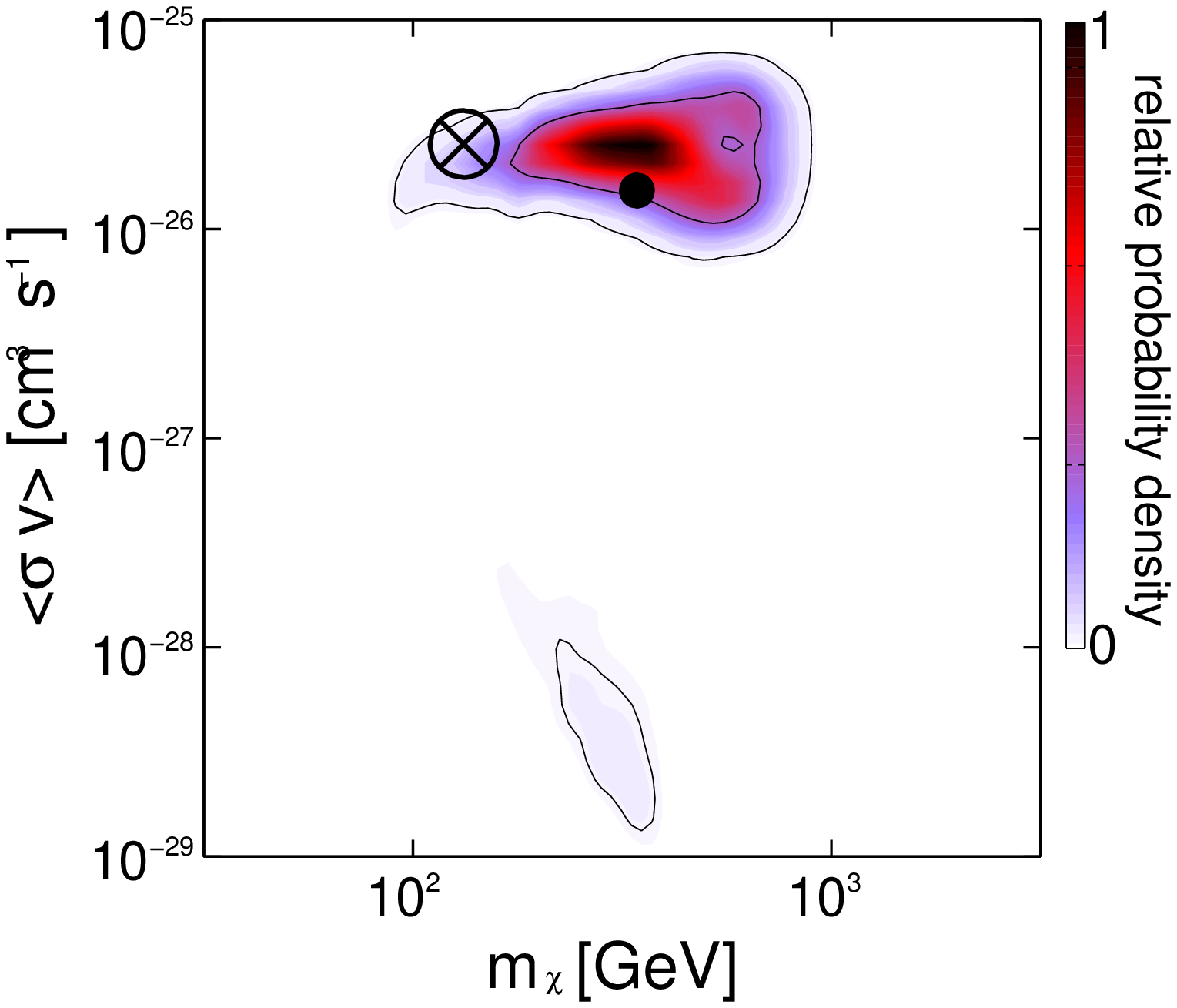}
\includegraphics[width=0.31\linewidth]{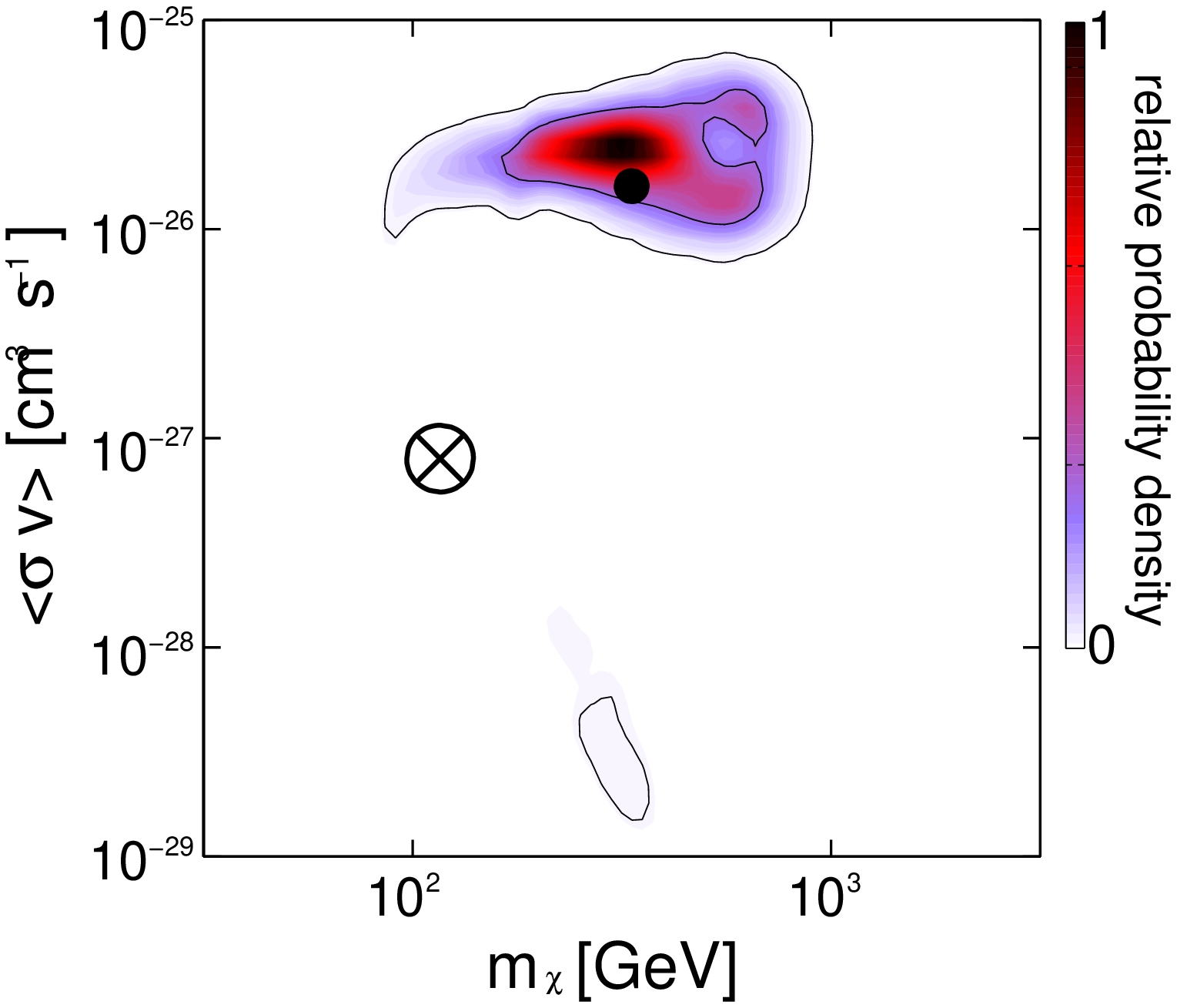}
\end{center}
\caption{Profile likelihood ratios
  $\mathcal{L}/\mathcal{L}_{\text{best fit}}$ (upper row) and
  posterior probability density functions (lower row) normalized to
  the best fit point in the $m_{\chi}$-$\langle \sigma v \rangle$
  plane. In the left column no H.E.S.S. data are included in the
  scans. In the middle column Carina and Sculptor data are included.
  In the right column SgrD is also included.}
\label{fig_dwarfs2}
\end{figure}

H.E.S.S. has observed the dwarf spheroidals Carina and Sculptor in
$2008$ and $2009$ for $14.8 \, \text{h}$ and $11.8 \, \text{h}$ live
time. No significant excess was detected \cite{bib_carscul}. We handle
these two sources in the same way as the SgrD. The original H.E.S.S.
paper gives event numbers, both in total and above some minimal energy
$E_{\text{min}}$, the resulting upper limits on the number of excess
events, and an integrated flux with $E>E_{\text{min}}$. In order to
estimate the flux as we did with the SgrD, we used a simple toy Monte
Carlo simulation to determine all combinations of event numbers with
$E>E_{\text{min}}$ that reproduce the given upper limits on the excess
of events. For our corresponding flux estimates, we then selected the
largest event numbers that delivered the stated upper limits, as
$E_{\text{min}}$ is chosen very close to the energy threshold of the
observations, so the majority of events should have
$E>E_{\text{min}}$. Because in the majority of combinations the
resulting estimated flux varies well within one standard deviation,
the error we make with this method is small. The estimated fluxes are
$\Phi (E > 320 \, \text{GeV}) = (-1.99 \pm 1.88) \cdot 10^-13 \,
\text{cm}^{-2} \, \text{s}^{-1}$ for Carina and $\Phi (E > 220 \,
\text{GeV}) = (0.42 \pm 3.38) \cdot 10^{-13} \, \text{cm}^{-2} \,
\text{s}^{-1}$ for Sculptor.  The resulting implications for scans of
the CMSSM parameter space can be seen in Figs.~\ref{fig_dwarfs1} and
\ref{fig_dwarfs2}.  In the centre panels of these figures, we see that
the addition of Carina and Sculptor -- with median values of the $J$
factors reported in \cite{bib_carscul}: $\bar{J}(\Delta \Omega)\Delta
\Omega \lvert_{\text{Carina}} = 1.35 \cdot 10^{-4} \, \text{sr}$ and
$\bar{J}(\Delta \Omega)\Delta \Omega \lvert_{\text{Sculptor}} = 1.91
\cdot 10^{-3} \, \text{sr}$ -- reduces the posterior probability of
the stau coannihilation region relative to the focus point.  This
effect is not so dramatic as was seen with the cored-profile SgrD in
the rightmost panels of Figs.~\ref{fig_sgd1} and \ref{fig_sgd2}.  In
the rightmost panels of Figs.~\ref{fig_dwarfs1} and \ref{fig_dwarfs2},
we also show the impact of including all three dwarfs, this time with
a SgrD $J$ factor calculated as the mean of the $J$ factors derived
from NFW and cored profiles of $\bar{J}(\Delta \Omega)\Delta \Omega
\lvert_{SgrD} = 0.327 \, \text{sr}$.  As expected, the coannihilation
region is further disfavoured by the inclusion of the SgrD, though
again not so severely as when this particular dwarf is employed with
the (maximal) $J$ factor corresponding to a cored density profile.
Profile likelihoods follow essentially similar trends to posteriors,
except for the fact that a highly isolated, very high likelihood
best-fit point has been found in the scan including only Carina and
Sculptor, but not in other scans.  When the profile likelihood ratio
is calculated using this best-fit value and plotted, the effect is to
make all parts of the parameter space appear to have low likelihoods
(i.e. essentially all of the allowed parameter space appears green in
the middle panels of Figs.~\ref{fig_dwarfs1} and \ref{fig_dwarfs2}).
This is easily understood as a result of the highly spiked nature of
the CMSSM parameter space; here the scan has in fact managed to find
its way part-way up the isolated focus point likelihood spike
identified in \cite{bib_Yashar}.  This spike is typically missed in
scans employing the standard configuration of the MultiNest algorithm
(as we do here) \cite{bib_Yashar, bib_SBproflike}, as this mode is
optimised more for mapping the posterior than producing
fully-converged profile likelihoods.  Posteriors produced with these
scanning parameters are of course fully converged; the profile
likelihood results we present here should therefore be taken with
something of a grain of salt, and the posteriors considered to be the
primary result of this paper.

\section{CMSSM likelihood scan with observations on the galactic halo} 
\label{sec_halo} 
 
\begin{figure}[ht] 
\begin{center} 
\includegraphics[width=0.31\linewidth]{noHess_2D_profl_1.eps} 
\includegraphics[width=0.31\linewidth]{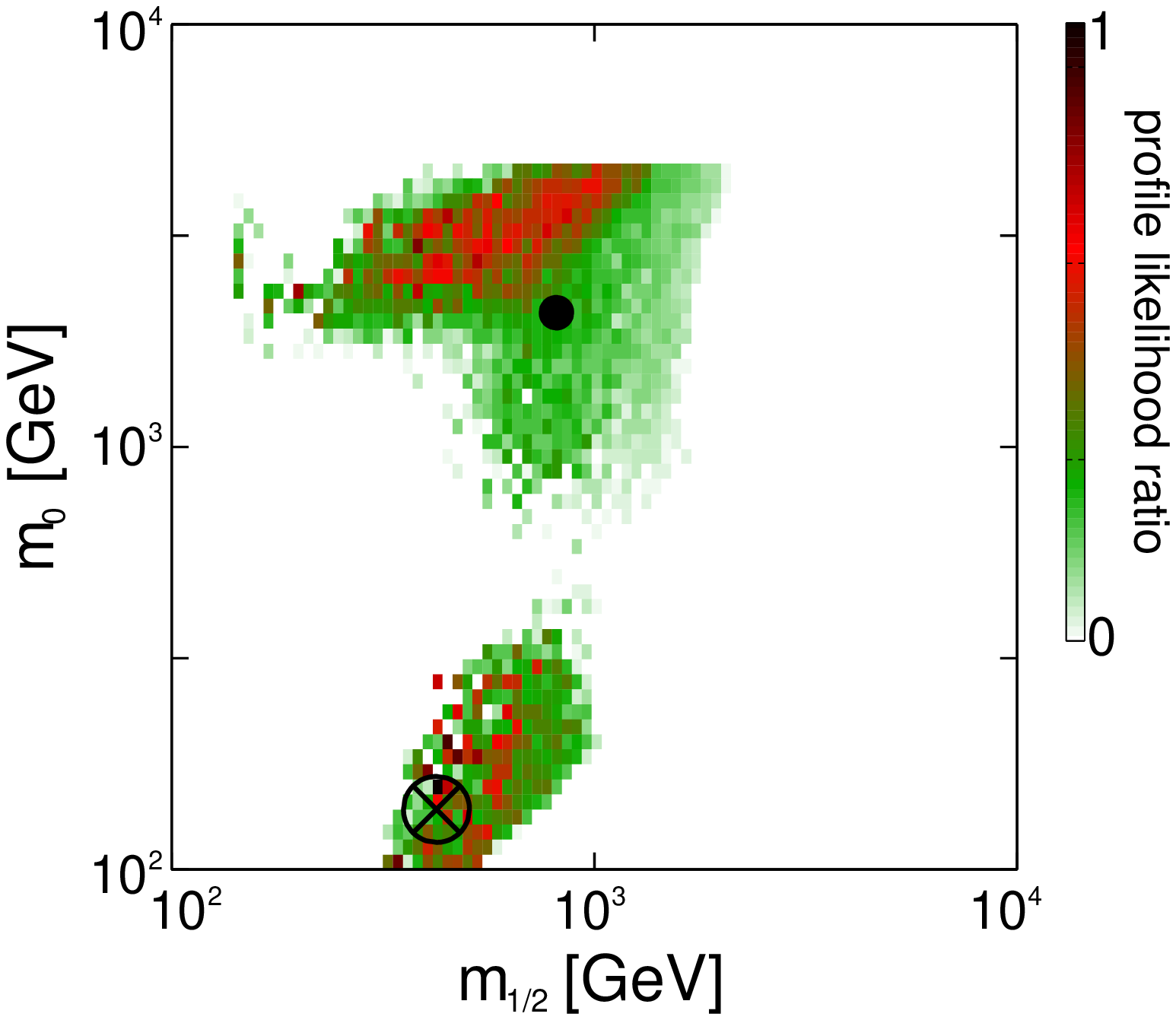} 
\includegraphics[width=0.31\linewidth]{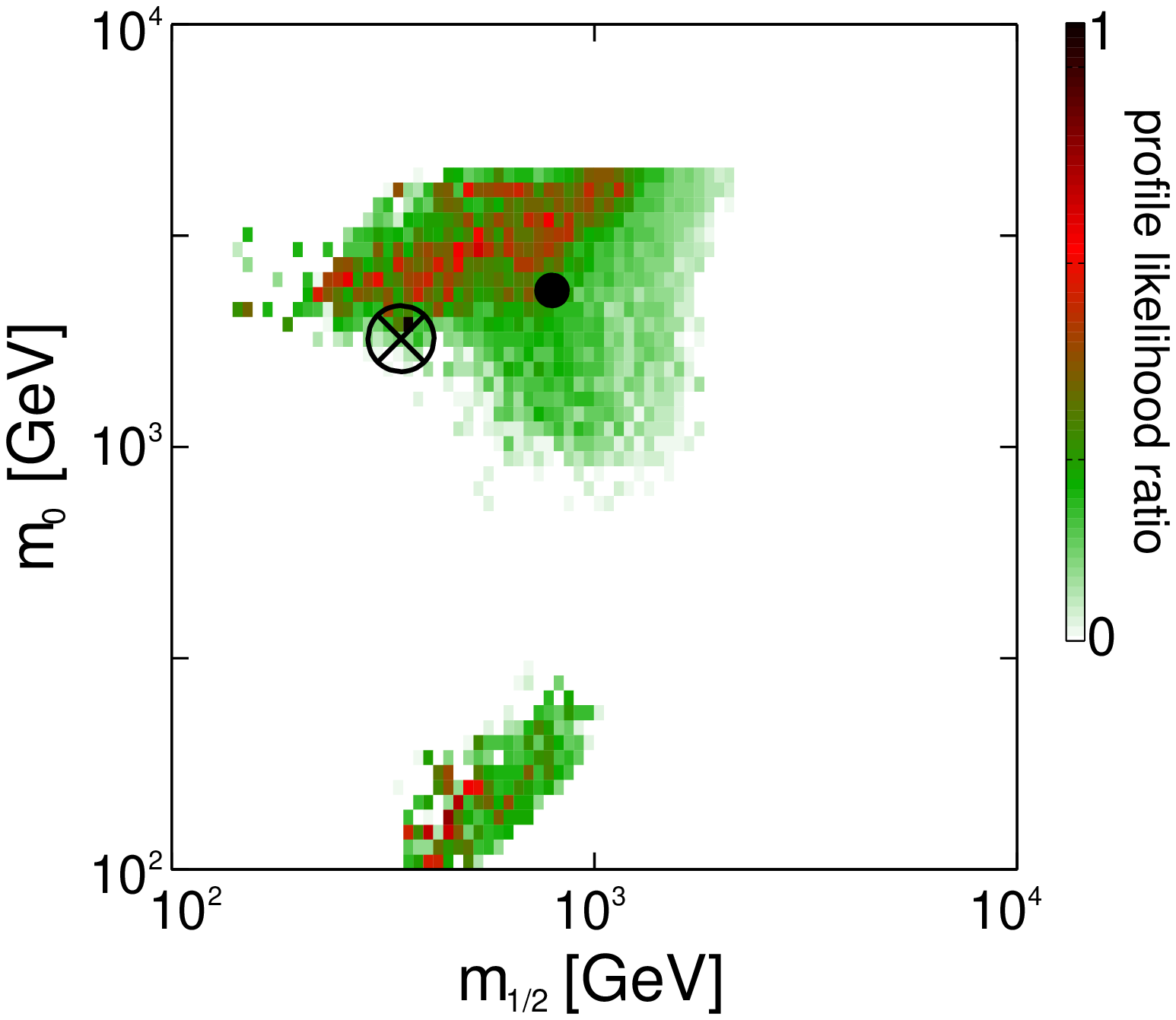} 
\includegraphics[width=0.31\linewidth]{noHess_2D_marg_1.eps} 
\includegraphics[width=0.31\linewidth]{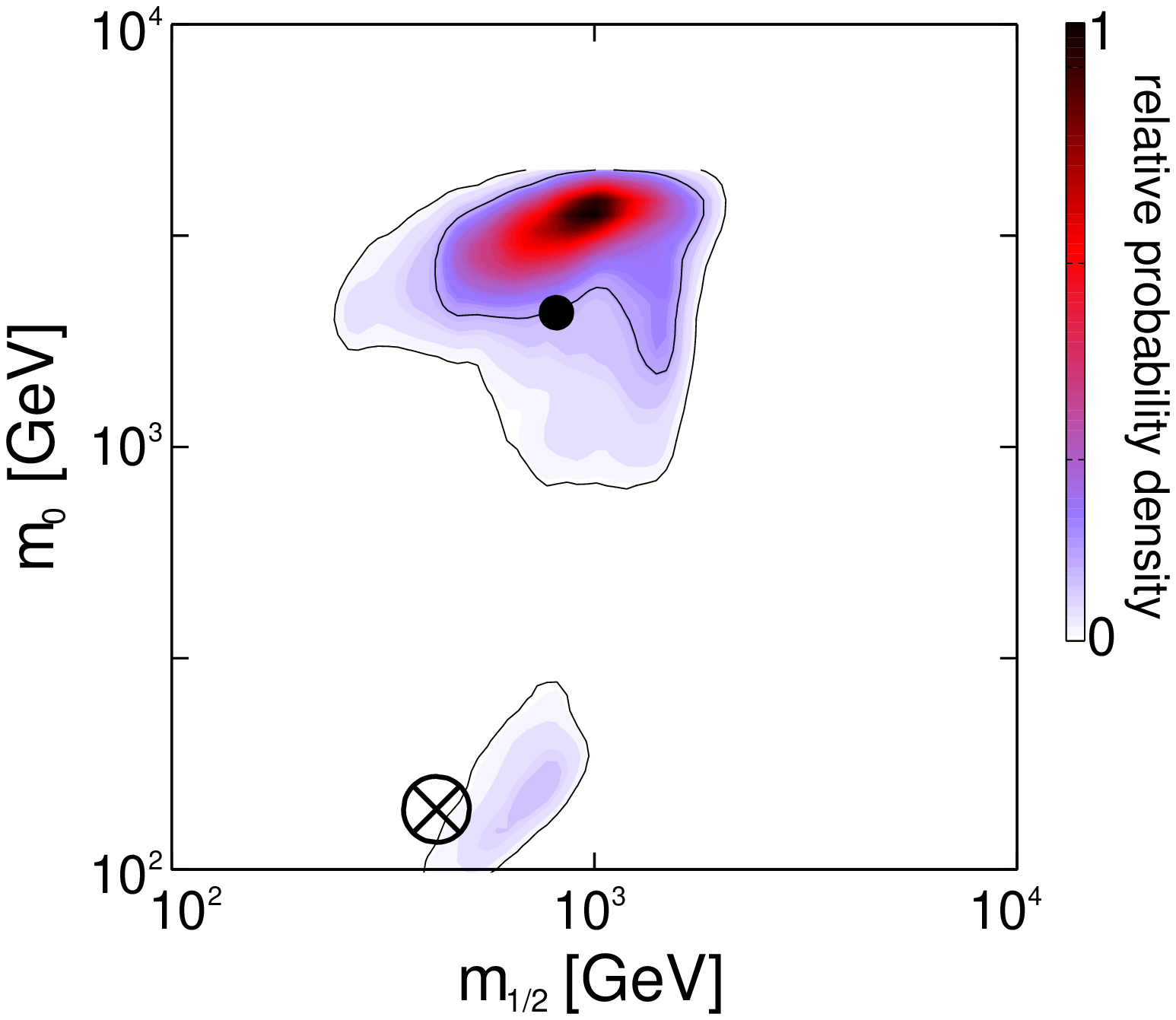} 
\includegraphics[width=0.31\linewidth]{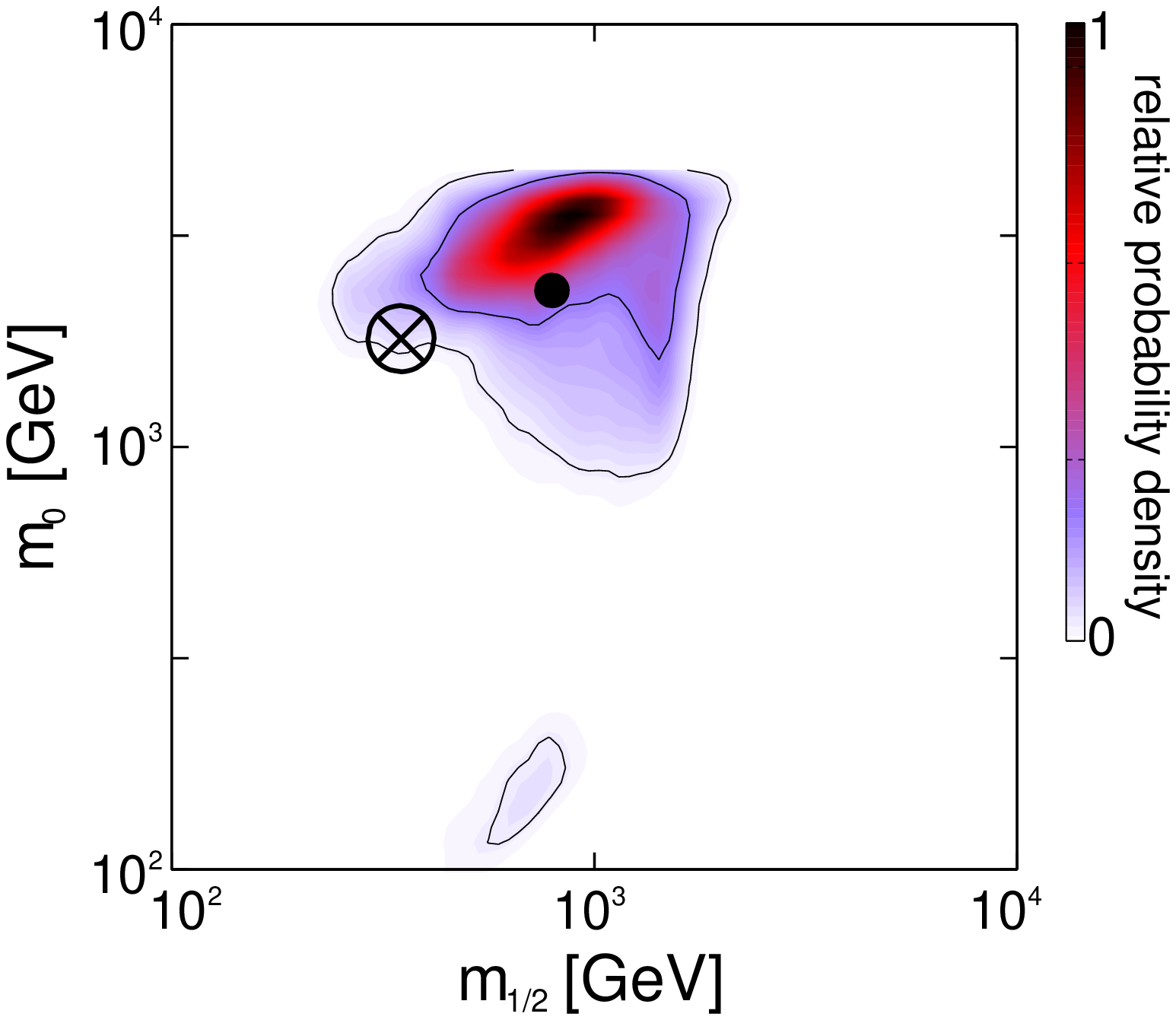} 
\end{center} 
\caption{Profile likelihood ratios 
  $\mathcal{L}/\mathcal{L}_{\text{best fit}}$ (upper row) and 
  posterior probability density functions (lower row) normalized to 
  the best fit point in the $m_\frac12$-$m_{0}$ plane. In the left 
  column no H.E.S.S. data are included in the scans. In the other two 
  columns observations on the galactic halo are included, together 
  with Carina and Sculptor (middle) and with Carina, Sculptor and the SgrD 
  (right).} 
\label{fig_halo1} 
\end{figure} 
 
\begin{figure}[ht] 
\begin{center} 
\includegraphics[width=0.31\linewidth]{noHess_2D_profl_7.eps} 
\includegraphics[width=0.31\linewidth]{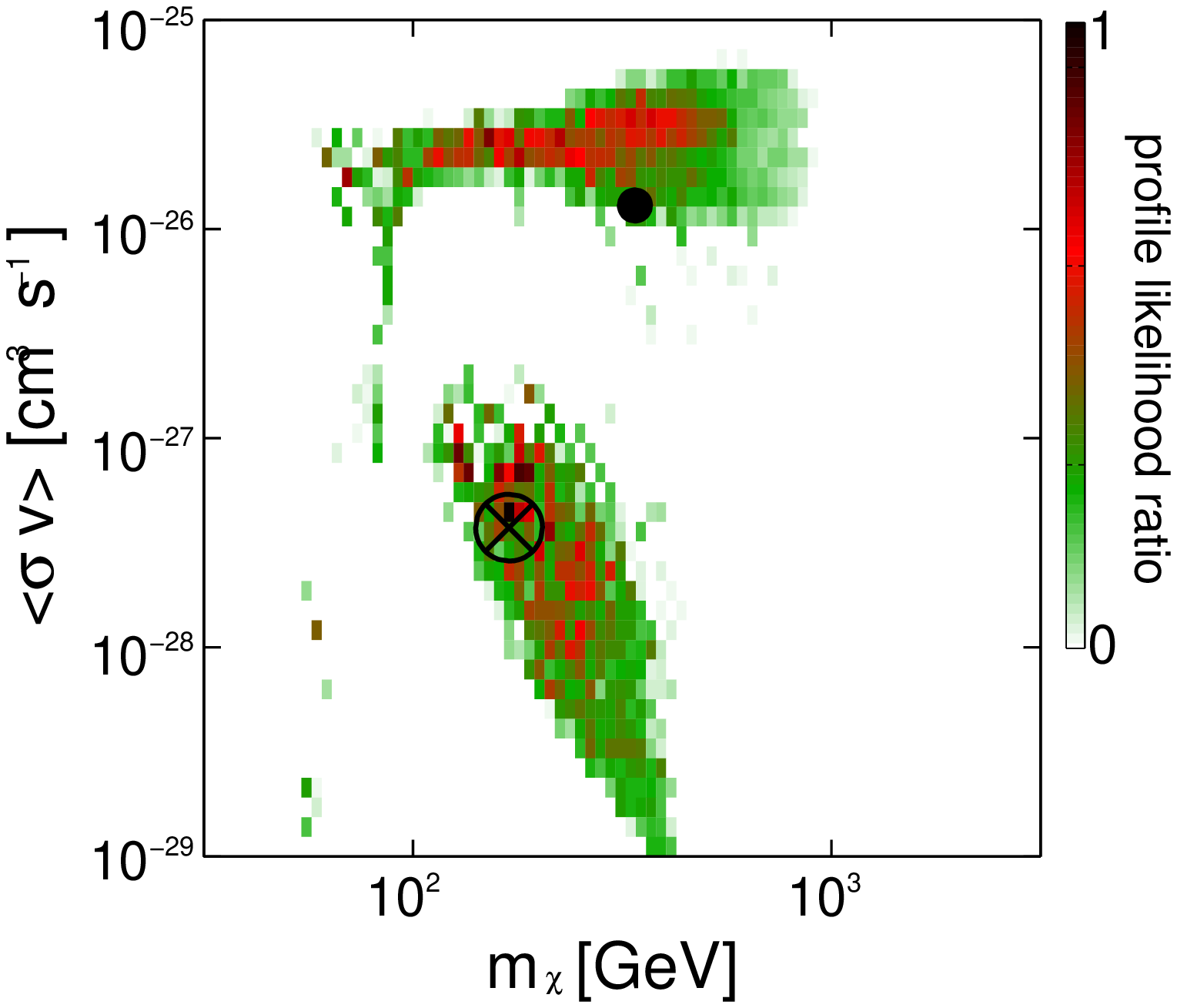} 
\includegraphics[width=0.31\linewidth]{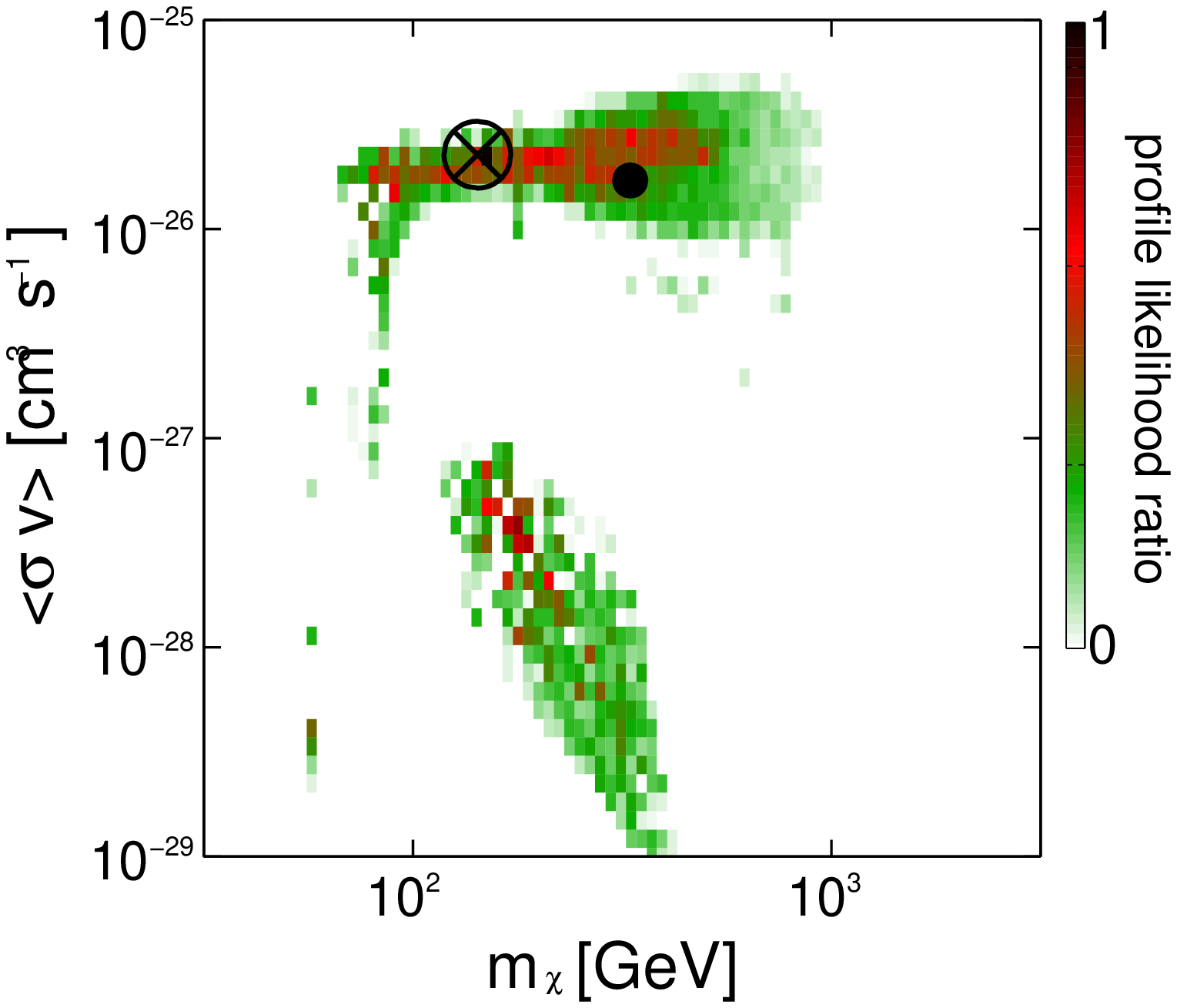} 
\includegraphics[width=0.31\linewidth]{noHess_2D_marg_7.eps} 
\includegraphics[width=0.31\linewidth]{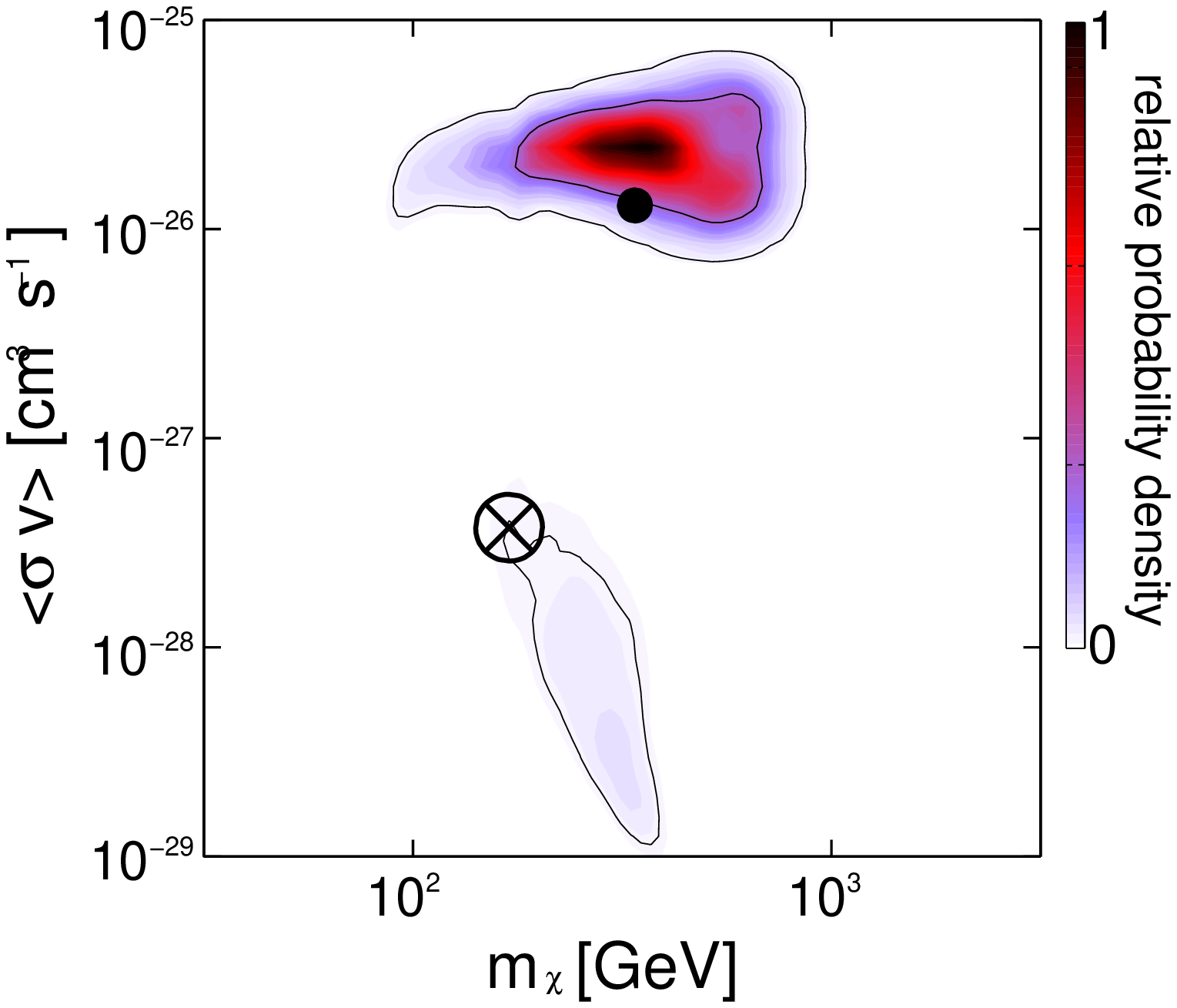} 
\includegraphics[width=0.31\linewidth]{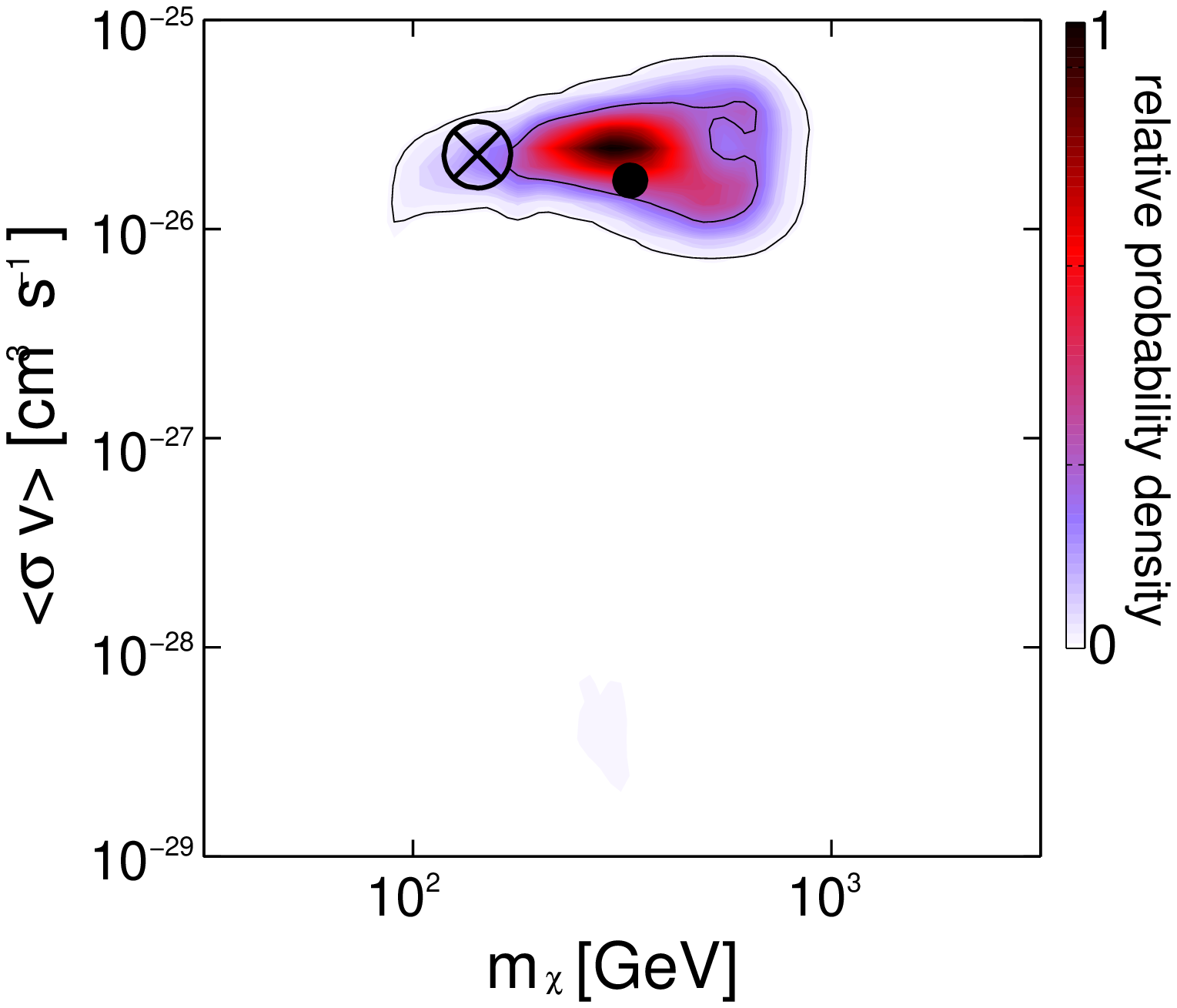} 
\end{center} 
\caption{Profile likelihood ratios 
  $\mathcal{L}/\mathcal{L}_{\text{best fit}}$ (upper row) and 
  posterior probability density functions (lower row) normalized to 
  the best fit point in the$m_{\chi}$-$\langle \sigma v \rangle$ 
  plane. In the left column no H.E.S.S. data are included in the 
  scans. In the other two columns observations on the galactic halo 
  are included, together with Carina and Sculptor (middle) and with  
  Carina, Sculptor and the SgrD (right).} 
\label{fig_halo2} 
\end{figure} 
 
H.E.S.S. has performed observations near the galactic centre in the
years from $2004$ to $2008$ in order to measure diffuse
$\gamma$-radiation from the galactic halo. The residual spectrum does
not show evidence of any excess $\gamma$-radiation \cite{bib_halo}.
This spectrum can be handled like the spectrum of the Galactic Centre,
with the only difference being that the observable is intensity rather
than flux.  As in the previous section, we assume for the halo a
median $J$ factor between the minimum and maximum values given by
\cite{bib_halo} $\bar{J}(\Delta \Omega) \lvert_{\text{halo}} = 1257$.
For scans including the halo, we also included Carina and Sculptor
(middle and rightmost panels of Figs.~\ref{fig_halo1} and
\ref{fig_halo2}), as well as the SgrD (rightmost panels of
Figs.~\ref{fig_halo1} and \ref{fig_halo2}).  Comparing the middle
panels of Figs.~\ref{fig_halo1} and \ref{fig_halo2} to the middle
panels of Figs.~\ref{fig_dwarfs1} and \ref{fig_dwarfs2}, we see that
the addition of the Galactic halo data to scans including Carina and
Sculptor in fact serves to \textit{increase} the relative probability
of the coannihilation region with respect to the focus point.  This is
because the halo constraint is rather weak, and serves only to
directly constrain a few focus-point models with large cross-sections,
slightly disfavouring the focus point in comparison to the
coannihilation region, and therefore tempering the negative effect of
the Carina and Sculptor dwarfs upon the relative probability of the
coannihilation strip.  When the SgrD is added (rightmost panels of
Figs.~\ref{fig_halo1} and \ref{fig_halo2}), this effect is essentially
swamped by the strong constraining effect of the SgrD data.  This
results in essentially the same level of preference for the focus
point in scans including the SgrD with or without the halo data;
differences between the rightmost panels of Figs.~\ref{fig_halo1} and
\ref{fig_halo2} and Figs.~\ref{fig_dwarfs1} and \ref{fig_dwarfs2} is
within the level of scanning noise.

\section{Summary, conclusions, and outlook} \label{sec_sum} We have
performed a scan over the CMSSM parameter space, taking into account a
large range of experimental data at the composite likelihood level,
and using nested sampling. We have done this in order to check what
constraints are placed on CMSSM models by the combination of H.E.S.S.
observations of dwarf spheroidal galaxies, the Galactic halo and the
Galactic Centre.

Due to the strong astrophysical $\gamma$-ray source in or very near
the GC, the search for DM there is strongly handicapped, so the data
are not very constraining. With unrealistic assumptions about the DM
density profile around the GC, we showed some example constraints on
the coannihilation region and focus-point neutralinos with large
masses. These examples show how the scanning technique will be useful
for future observations with the next generation of $\gamma$-ray
experiments, such as CTA.

For dwarf galaxies and the Galactic halo we also obtained constraints
on the coannihilation region and high-mass parts of the focus point,
even with realistic density profiles.  These constraints result from
the combination of the energy reach of H.E.S.S. and a full treatment
of IB.  Our results give the tightest constraints to date upon the
coannihilation region of the MSSM.

There are however still large uncertainties in the DM density profile
of the SgrD, due to strong tidal forces \cite{bib_hessSGD,bib_Viana}.
This is unfortunate, as the SgrD potentially provides the strongest
constraint on CMSSM coannihilation models.  Future scans and limits
based on the SgrD should become more robust as they eventually come to
include the the dark matter halo parameters as nuisances, and
observational constraints upon those parameters improve.  Ultimately
however, we see that including observations of Carina and Sculptor
along with those of the SgrD, and assuming median values of all $J$
factors, results in almost as strong a constraint on the
coannihilation region as taking just the SgrD on its own, and using a
maximal $J$ factor.  This speaks strongly to the robustness of the
results we have presented in this paper.

The recently presented results from LHC \cite{bib_Aad:2011hh} are not
directly comparable with our results, since constraints are presented
for fixed $\tan \beta=3$ and $A_{0}=0$, which is actually not part of
the most favoured $68\%$ region that we find. However, ATLAS
constrains gaugino masses below about $310 \, \text{GeV}$ and scalar
masses below about $740 \, \text{GeV}$. Most of the favoured region
that we find is at either larger gaugino or larger scalar masses.
Thus, the present ATLAS constraint can be expected to have a minor
effect on the results presented here, see e.g.
\cite{bib_Bertone:2011nj} for a more detailed discussion.

\section{Acknowledgments}
We would like to thank Agnieszka Jacholkowski and Ullrich Schwanke for
fruitful discussions. We are grateful to the Swedish Research Council
(VR) for financial support. JR is grateful to the Knut and Alice
Wallenberg Foundation for financial support. JC is a Royal Swedish
Academy of Sciences Research Fellow supported by a grant from the Knut
and Alice Wallenberg Foundation. PS is supported by the Lorne Trottier
Chair in Astrophysics and an Institute for Particle Physics Theory
Fellowship.

\end{document}